\documentclass[sigconf]{acmart}
\usepackage{array} %cannot?
\usepackage{longtable} %cannot?
\usepackage{multirow}
\usepackage{float}
\usepackage{tabularx} %cannot?

\usepackage[normalem]{ulem}
\AtBeginDocument{%
  }

\setcopyright{acmlicensed}
\copyrightyear{2026}
\acmYear{2026}
\setcopyright{cc}
\setcctype{by}
\acmConference[CHI '26]{Proceedings of the 2026 CHI Conference on Human Factors in Computing Systems}{April 13--17, 2026}{Barcelona, Spain}
\acmBooktitle{Proceedings of the 2026 CHI Conference on Human Factors in Computing Systems (CHI '26), April 13--17, 2026, Barcelona, Spain}
\acmPrice{}
\acmDOI{10.1145/3772318.3790657}
\acmISBN{979-8-4007-2278-3/2026/04}

\begin{document}
\newcommand{\artefact}{70 } 

\title{Entangled Life and Code: A Computational Design Taxonomy for Synergistic Bio-Digital Systems}
\titlenote{Inspired by \textit{Entangled Life} by M. Sheldrake \cite{sheldrake_entangled_2020}}

% \zb{Symbiotic Systems, Living Logic, Living Patterns, Bio Binary, Binary Bio, Dual Digits, Blended Bits, Bio Bits, Biotic Bits }\ks{maybe something that doesn't explicitly use ``digital''-only constructs? Living Lexicon? or Beyond Binaries (/Booleans?)?} \zb{I like Beyond Binaries} \ek{A Taxonomy and Pathways for Designing Synergistic Bio-Digital Systems}
% \ek{Entangled Life and Code: A Design Taxonomy and Pathways for Synergistic Bio-Digital Systems} \zb{old: A Computational Design Taxonomy for Synergistic Bio-Digital Systems}

%\ek{Computational Design Taxonomy for Synergistic Bio-Digital Systems across Time and Scale
%or
%A Computational Design Taxonomy for Synergistic Bio–Digital Systems Toward Regenerative Ecologies; 
%or Synergistic Bio–Digital Systems Toward Regenerative Ecologies: A Computational Design Taxonomy
%}

\author{Zoë Breed}
\email{z.breed@tudelft.nl}
\orcid{0009-0001-0882-5969}
\affiliation{%
    \department{Knowledge and Intelligence Design}
  \institution{Delft University of Technology}
  \city{Delft}
  \country{Netherlands}
}
\author{Elvin Karana}
\email{e.karana@tudelft.nl}
\orcid{0000-0001-9598-2753}
\affiliation{%
    \department{Materializing Futures}
  \institution{Delft University of Technology}
  \city{Delft}
  \country{Netherlands}
}
\author{Alessandro Bozzon}
\email{a.bozzon@tudelft.nl}
\orcid{0000-0002-3300-2913}
\affiliation{%
\department{Knowledge and Intelligence Design}
  \institution{Delft University of Technology}
  \city{Delft}
  \country{Netherlands}
}
\author{Katherine W. Song}
\email{k.w.song@tudelft.nl}
\orcid{0000-0001-7580-0035}
\affiliation{%
\department{Knowledge and Intelligence Design}
  \institution{Delft University of Technology}
  \city{Delft}
  \country{Netherlands}
}

\renewcommand{\shortauthors}{Breed et al.}

\begin{abstract}
Bio-digital systems that merge microbial life with technology promise new modes of computation, combining biological adaptability with digital precision. Yet realizing this potential symbiotically -- where biological and digital agents co-adapt and co-process -- remains elusive, largely due to the absence of a shared vocabulary bridging biology and computing. Consequently, microbes are often constrained to uni-directional roles, functioning as sensors or actuators rather than as active, computational partners in bio-digital systems. In response, we propose a taxonomy and pathways that articulate and expand the roles of biological and digital entities for synergetic bio-digital computation. Using this taxonomy, we analysed \artefact systems across HCI, design, and engineering, identifying how biological mechanisms can be mapped onto computational abstractions. We argue that such mappings enable computationally actionable directions that foster richer and reciprocal relationships in bio-digital systems, supporting regenerative ecologies across time and scale while inspiring new paradigms for computation in HCI.
  
\end{abstract}



\begin{teaserfigure}
  \centering\includegraphics[width=0.69\textwidth]{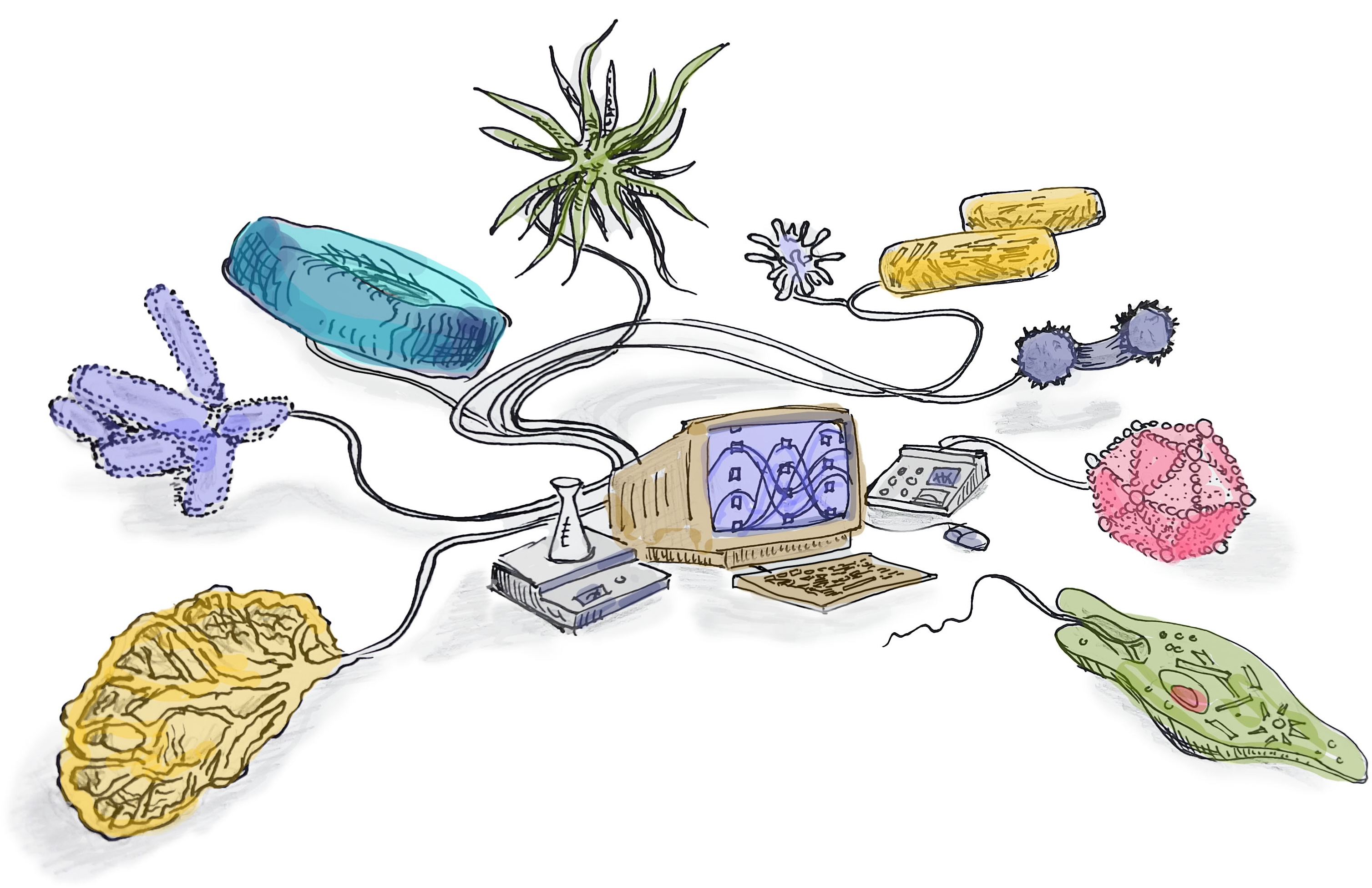}
  \caption{Illustrative sketch depicting microorganisms approaching digital components from all sides, merging together within the visualization platform introduced in this paper.}
  \Description{Drawing of nine microorganisms approaching from all sides, connected by lines to a central computer. On the computer screen, the visualization platform introduced in this paper is displayed.}
  \label{fig:teaser}
\end{teaserfigure}

%%
%% Process the author and affiliation and title
%% information and build the first part of the formatted document.
\maketitle

\section{Introduction}
Admiration for the beauty, complexity, and intelligence of the living world around us has captured the hearts and minds of humans since the beginning of our existence \cite{edward_o_wilson_biophilia_1986}. Within Human-Computer Interaction (HCI), regenerative design approaches that forefront care for the natural world and ecological restoration have emerged through several burgeoning sub-communities and themes, such as biodesign \cite{gough_nature_2020}, biological HCI \cite{pataranutaporn_biological_2018}, more-than-human-centred design \cite{wakkary_things_2021}, ecological HCI \cite{lu_ecological_2024}, and regenerative material ecologies \cite{nicenboim_regenerative_2025}. These communities have generated theories, artefacts, and frameworks that guide how we should incorporate and respect living organisms when designing new technologies. Nonetheless, a gap persists between these communities and mainstream areas of HCI, which constantly seek more responsive and more ``intelligent'' computational systems. For computational systems architects, the fast, digital, and deterministic modern computing paradigms appear to be incompatible with living organisms, which instead are often seen as slow, analogue, and probabilistic. 

Bridging this gap is imperative from multiple perspectives. First, in pursuit of Weiser's vision of ubiquitous computing in which technology seamlessly blends into the environment \cite{weiser_computer_1999}, it is no longer sufficient to innovate solely using artificial systems. Truly dissolving the boundary between the technological and the ecological demands literal, not just metaphorical, integration of elements of the world around us into computational systems. Additionally, the trajectory of modern intelligent interactive technologies is unsustainable in at least two aspects: materially, with electronic waste rapidly accumulating in landfills \cite{cornelis_p_balde_global_2024}, and energy-wise, with artificial intelligence (AI) models demanding excessive amounts of energy to train, test, and deploy \cite{inie_how_2025,luccioni_power_2024}. Replacing system components with the innate computational capabilities of biological materials and living organisms could open doors to materially and energetically sustainable interactive systems. Furthermore, from a regenerative perspective, it is crucial to articulate how to design synergistic bio-digital systems that foster ecological health and do not exploit living organisms as drop-in components. Bio-digital hybrid systems\footnote{We adopt the term ``bio-digital'' here from prior work in HCI. In this paper, we specifically mean ``bio-'' to describe systems incorporating living organisms (not only bio-derived). We use ``digital'' to encompass non-living electronics that can also have non-digital architectures (e.g., analogue).} provide a unique opportunity where living organisms and electronics can co-evolve in mutualistic relationships, where each component enhances and adapts to the other. Such dynamics open possibilities for artefacts\footnote{We understand ``systems'' (more common in engineering / HCI) and ``artefacts'' (more common in design) to be interchangeable in this paper.} that not only sustain but actively enable new forms of resilience and circularity -- repairing themselves, adapting to new conditions, and contributing to ecological cycles. Digital components can and should be active, ``living'' participants in regenerative ecologies, not just tools for productivity \cite{bedau_living_2010}. Still, for designers, translating biological affordances into computationally actionable abstractions is elusive. Consequently, many existing bio-digital systems, while aesthetically striking and conceptually provocative, harness only a small slice of the computational capabilities, and subsequently regenerative potential, of the living organisms that they incorporate. Meanwhile, computational systems designers vaguely sense the potential of living systems but often lack the deep understanding of biology needed to leverage these capabilities in practice. 

How, then, can we design bio-digital systems that symbiotically realize the potential of both components and translate the capabilities of living organisms to abstractions that are tractable for biodesigners and computational systems designers alike? We believe that developing a shared vocabulary is key to this translation, providing common ground for collaboration, reducing miscommunication, and enabling knowledge transfer across disciplines. Researchers in HCI and beyond have attempted to do this by mapping biological behaviours one-to-one with computational components, such as logic gates \cite{wang_engineering_2011,de_silva_molecular_2007}. Others have developed bottom-up, organism-centric frameworks \cite{lucibello_bio-digital_2024}. While the former approach has led to laudable prototypes, it force-fits biological organisms into paradigms developed for silicon-based computers, neglecting emergent biological modes of ``computation'' \cite{goni-moreno_angel_biocomputation_2024,grozinger_pathways_2019}. Meanwhile, the latter inspires novel concepts but lacks traction among computational systems designers, who depend on formal abstractions, hierarchies, and input-output specifications expressed in metrics like voltages and frequency. 
To date, no concerted effort has been made to develop a shared vocabulary for bio-digital computation. The computational systems focus is significant because interactive properties -- responsiveness, adaptability, sensing, and feedback -- are fundamentally enabled by the underlying computational architecture.
%\ale{Meta-comment: what we originally framed as a "symmetric" (bio-design vs. computational) is obviously going to lean in one direction - probably design? I wonder if at some point we need to be explicit about this "design drift", or if we should always try to balance this 2 perspectives. In the paragraph above, the balance is there. But also see my comment before. }\ks{I think it indeed naturally leans design, but it is nonetheless valuable to try to balance them throughout (or at least periodically)...w/ the new discussion section I think we do this}

\subsection{Contribution}
This paper contributes to HCI a computational design taxonomy and vocabulary for bio-digital interfaces that is both biologically faithful and computationally actionable. This vocabulary is intended to bridge different perspectives present among HCI researchers, including (bio)designers and electronics systems designers. % \ale{Shall we be explicit about who are the intented recipients of this effort? Designers? Bio-designers? HCI researchers? Computer Scientists? Later in Section 3 we state that "The purpose of our computational taxonomy is to provide biodesigners". At the very least we should be uniform}
We operationalize this as a scaffold to interpret \artefact existing bio-digital systems, with the goals of both describing the computational roles of biological and digital components and prescribing novel computational partnerships. %\ale{In Section 4.1 there are two very interesting  complementary questions. Shall we mention the goals behind the questions here? Our goal is now only captured by the last sentence in 1.1 }\ks{killed the q's because 1 was answered in results; the other in the discussion...but looped their spirits in here}
Secondarily, we contribute an open-source database, pre-populated with our analysed subset of bio-digital systems, and an interactive visualization platform. We use the platform to systematically explore relationships between organisms and digital components across different temporal and spatial scales within a computationally tractable framework. We demonstrate how such a structured exploration reveals existing patterns and missed opportunities for bio-digital systems that are grounded in feasible technical implementations.
\newpage
\subsection{Researcher-Designer Approach}
We adopt a first-person, reflexive designer-research approach \cite{chen_crafting_2021}, drawing on our own multi-disciplinary knowledge and perspectives to propose a taxonomy. We are a team of computer scientists (two PhDs with a collective 25 years publishing at computer science and HCI venues) and biodesigners (one PhD and one PhD candidate with a collective 25 years publishing at (bio)design and HCI venues). While we are employed at the same academic institution, our academic trajectories span across four different countries in North America, Europe, and the Middle East. Our geographic and academic backgrounds shape (and bias) our perspectives on computation, cognition, intelligence, and nature. This is our first collaborative effort, positioning this paper as both a theoretical contribution and an open, Research-through-Design-inspired methodology that demonstrates the type of cross-disciplinary pollination needed to develop holistic bio-digital systems.

\section{Related Work}
\subsection{Regenerative Systems in HCI}
The notion of regeneration -- of designing not only to sustain but also to restore and enrich ecosystems -- has been rapidly gaining traction within the HCI and design communities. Regenerative design moves beyond conventional sustainability paradigms, challenging designers to not only minimize the use of non-degradable materials but also contribute positively to the vitality of living organisms across material, social, and ecological dimensions \cite{lyle_regenerative_1994,wahl_designing_2016,gibbons_regenerativenew_2020}. Recognizing that anthropocentric design has contributed to ecological degradation, regenerative design draws from other scholarly movements, such as posthumanism \cite{haraway_when_2007, bellacasa_matters_2017, braidotti_theoretical_2019}, that de-centre human needs and acknowledge the agency of non-human entities \cite{wakkary_things_2021,oogjes_repertoires_2022,sharma_post-growth_2023,nicenboim_designing-ai_2024,nicenboim_decentering_2025,giaccardi_makings_2025}. Still, implementing regenerative approaches is complex in practice. %\ek{>> here talk about how the inherently toxic nature/slicon-based computation.. and limited understanding of organisms/ecosystems around us. }
For one, we lack ``ecoliteracy'' and an understanding of the ecosystems and living worlds around us \cite{toner_integrating_2023}. Moreover, the existing digital infrastructure and tools that we have come to depend on rely on manufacturing processes and materials that are inherently energy-intensive, toxic, and non-renewable \cite{wang_environmental_2023,cornelis_p_balde_global_2024,luccioni_power_2024,inie_how_2025}. 

To that end, the mindful creation of ``living artefacts'' -- tangible systems integrating living organisms -- introduces the inherent capacities of life, such as growth, repair, and transformation, into material practices. These systems offer novel functionalities, aesthetics, opportunities for mutualistic care, and re-conceptualizations of habitats \cite{karana_living_2020}, thereby enabling them to engage in ecological cycles and support regeneration  \cite{karana_living_2023}. At the same time, these systems have the potential to cultivate sensibilities and ecological literacies, fostering deeper awareness and care in shared habitats \cite{karana_living_2023, zhou_living_2024}. At CHI'25, the panel on Regenerative Material Ecologies in HCI \cite{nicenboim_regenerative_2025} articulated the ambition of regenerative systems while identifying a critical challenge and missed opportunity when it comes to actually designing with living organisms in a way that is regenerative: while regenerative design philosophy offers guidance for why and what we might design, the ``how'' remains underdeveloped, with unclear pathways for implementation and especially integration with digital infrastructure and tools.

\subsection{Bio-Digital Interfaces in HCI}
Bio-HCI, the subject of several workshops at CHI and DIS \cite{gough_nature_2020,forman_living_2023,zhu_wearable_2025,kim_microbe-hci_2021}, is a term encompassing approaches to integrate biomaterials -- both living and derived from living systems -- into interactive interfaces \cite{bell_integrating_2025}. While bio-HCI is not synonymous with regenerative design, projects in this space showcase the potential for living artefacts that can grow, adapt, and respond to their environments in ways that conventional materials cannot. Many transform garments, surfaces, and other materials into habitats for living organisms, endowing otherwise passive products with novel aesthetics, functionality, and responsiveness \cite{yao_biologic_2015,oxman_dermi-domus_2017,zhu_livingloom_2025,groutars_habitabilities_2025}. Others enable direct interactions between humans and organisms without digital mediation, revealing new paradigms of care-centric interactions, making, and ecological awareness \cite{ofer_designing_2021,bell_reclaym_2022,bell_me_2023,bell_bio-digital_2024,zhou_living_2024,risseeuw_reactivate_2024}.

Combinations of digital components with living systems have opened up rich possibilities for HCI, contributing, for example: digital systems with living substrates or scaffolds \cite{krieger_living_2025,zhu_livingloom_2025}, augmented fermentation systems \cite{chen_nukabot_2021,bell_bio-digital_2024}, living displays in which digital components activate visible organism responses \cite{barati_living_2021,breed_algae_2024}, systems that sonify living processes \cite{riggs_mold_2025, miranda_composing_2018}, biotic games in which humans control living microorganisms through a digital interface \cite{kim_interactive_2016, kim_new_2018, lam_pac-euglena_2020, riedel-kruse_design_2010, guo_slimo_2025}, interactive habitats and observation platforms \cite{groutars_flavorium_2022}, digital representations of living organisms through digital twinning and/or extended reality interfaces \cite{hu_immersive_2024,vu_addressing_2024}, and hybrid bio-digital fabrication platforms \cite{iwasaki_silk_2017,smith_hybrid_2020, postl_mold_2024}. Plants have especially been broadly studied in bio-digital systems within HCI \cite{kuribayashi_io_2007,poupyrev_botanicus_2012,manzella_plants_2013,sareen_cyborg_2019,chang_patterns_2022,fell_biocentric_2022,hu_immersive_2024,hu_designing_2024,zhu_livingloom_2025,luo_plantmate_2025}. Such studies showcase novel interactions and technical feasibility. Still, to move beyond one-off implementations, we need design frameworks that help articulate and inspire how biological and digital components can support one another, especially towards regenerative outcomes.

Existing HCI frameworks provide a valuable starting point. Zhou et al.'s taxonomy of digital tools for biodesign takes an organism-centric approach, elucidating how technology can play the roles of understanding, embodying, and perpetuating habitats for living organisms \cite{zhou_habitabilities_2022}. Kim et. al further elaborate upon the role of digital components as a surfacing mechanism \cite{kim_surfacing_2023}. On the more mechanistic side, Pataranutaporn et al.'s \textit{Living Bits} framework proposes a set of computational characteristics -- input, process, output, language, scale, speed, power, and recyclability -- that structures comparisons between biological and digital systems and guides future bio-digital hybrid system design \cite{pataranutaporn_living_2020}. \textit{Living Bits} is a significant first step in foregrounding computation as a shared lens for bio-digital integration, offering a high-level classification for certain dimensions. However, much work remains to unpack some aspects further: in particular, ``processing'' is treated as a single, undifferentiated function, obscuring the diverse and fine-grained computational roles that living systems might play.
%\ek{Can we phrase this to acknowledge that it serves as a good entry point and offers a high-level classification—one of the motivations for our paper was precisely to unpack some of these high-level aspects, such as processing, which is currently… .. just to give it a positive tone} The \textit{Living Bits} framework provides a starting point in this regard but treats computational ``processing'' monolithically. 
\subsection{Frameworks and Examples for Computing with Living Organisms Beyond HCI}
%Despite growing excitement in bio-digital interfaces, hybrid systems that effectively utilize advantages of both biological organisms and silicon electronics, especially toward regenerative outcomes, remain elusive. To address this gap, 
Bio-digital frameworks in HCI elucidate how digital systems can support living organisms, maintaining or creating habitats, modelling and surfacing organisms' behaviours, or co-fabricating materials together. An underlying assumption is that the computation is done by the digital component; after all, digital systems are built to compute. Critically, however, organisms compute too, and designing bio-digital systems without unpacking the computational dimension misses core aspects of both digital and biological systems' existence. Accordingly, theoretical computer scientists have proposed formal abstractions to understand how biological computation compares with other forms of computation. Jaeger et al.'s generalized computational framework describes how models of computation, input/output data, and environmental factors translate between theoretical and physical systems, agnostic of substrate, including biological ones \cite{jaeger_exploring_2021}. Stepney et al. define formal criteria for when biological processes constitute ``genuine'' computation \cite{stepney_journeys_2005,horsman_abstraction_2017,stepney_co-designing_2019}. Păun and other ``natural computing'' researchers have developed theoretical models such as membrane computing, which abstracts the hierarchical structure and biochemical processes of biological cells into formal, generalizable computing frameworks \cite{paun_membrane_2002}. While theoretically rigorous, such frameworks operate at high abstraction levels that are not readily translatable for practitioners building physical bio-digital systems.

Engineering design researchers have tackled the biological-digital gap by focusing on language barriers between disciplines. Several have proposed systematic approaches for identifying keywords from biology literature and translating these into meaningful analogies for engineering design (e.g., the biological keyword ``lyse'' has functional correspondents such as ``end'' or ``interrupt'') \cite{cheong_biologically_2011,chiu_biomimetic_2007}. Such approaches are still fairly abstract and primarily support biomimetic or bio-inspired design.

On the other end, roboticists developing bio-hybrid ``cyborgs'' have proposed mechanistic frameworks that specify mechanical analogues for biological components. Ricotti and Menciassi's framework outlines traditional mechatronic parts as targets for biological component development \cite{ricotti_bio-hybrid_2012}. Extending beyond individual robots and mechanical parts, Romano et al. address social dimensions of bio-hybrid systems, developing control protocols for animal-robot and multi-organism coordination in swarms \cite{romano_unveiling_2021}. These approaches outline future avenues of mechanical integration and communication for bio-digital robots but do not readily generalize to non-robotic systems.

There are yet other frameworks for bio-digital systems emerging from investigations centred on specific organisms or specific computational paradigms. Slime mould (\textit{Physarum polycephalum}) in particular has been extensively characterized by Adamatzky and other ``unconventional computing'' researchers, who have implemented slime mould analogues to digital components, like logic gates and memristors, and have proposed theories for how \textit{Physarum} ``computes'' \cite{adamatzky_physarum_2007,adamatzky_physarum_2010,adamatzky_advances_2016,bonifaci_physarum_2012}. Very recently, reservoir computing, a lightweight machine learning model, has been applied to capture fungal mycelia growth and bacterial colony dynamics as computation \cite{tompris_mycelium_2025,telhan_morphologically_2025,ahavi_cellular_2025}. These cross-disciplinary efforts reflect the growing recognition of biological systems' computational potentials and the diversity of approaches for harnessing them.

\subsection{Summary of Accounts}
These existing works provide valuable foundations for bio-digital system design in HCI, contributing important perspectives on theoretical models (computer science), conceptual translation and analogy-building (engineering design), organism- and system- specific mechanical integration (multiple disciplines), and interaction paradigms (HCI). Yet none provide a holistic vocabulary for articulating the computational functions that living organisms and digital components can perform for one another in hybrid systems. %\new{As interaction unfolds through computational processes -- how systems detect, transform, compare, and respond to information -- articulating the relationship between the technical functions and interactional properties they give rise to is needed to align technical processes with meaningful user interaction.}% \ks{I put something like your previous version in the intro...see what you think of this rev here instead, to avoid repetition} \zb{Yes, better there}
While existing frameworks characterize interaction modalities and design roles, they offer limited guidance for implementing these interactions. This requires understanding the underlying computational capacities of system components. Establishing this understanding is an essential step if such systems are to move beyond interaction and toward regeneration, where biological and digital components form mutually beneficial partnerships that support larger ecosystems.

% \zb{The idea was to place the interaction + technical functions sentence here, but what do you think maybe 2.3 is better.  }\ks{I had the same thought just now -- put it in 2.3, but then in the end i think this is better?}
\section{Computational Design Taxonomy} \label{sec:taxonomy}
%\ek{It would be helpful to provide the reader with a brief roadmap of what follows from here.> first, we introduce a preliminary taxonomy based on […]; next, we examine […]; offer a tool ... and finally, we discuss […]. }
In this section, we introduce a computational design taxonomy for bio-digital systems, drawing upon existing multi-disciplinary computational frameworks. In subsequent sections, we use this taxonomy to examine patterns in current microorganism-based bio-digital systems. We present an interactive visualization platform and open-source database to facilitate exploration and analysis of bio-digital systems using our taxonomy as a scaffold. Finally, we discuss how our analysis generates concrete opportunities for bio-digital systems, surfaces new considerations for developing regenerative systems, parallels similar considerations in other sub-communities of HCI, and suggests future implications for the taxonomy itself.

\subsection{Formulation}

The purpose of our computational taxonomy is to provide (bio)\-designers and HCI researchers with a classification and vocabulary of essential computational functions in bio-digital systems at an abstraction level that is actionable -- appropriately grounded in computational theories with clear physical implementation implications -- while avoiding prescription to a single architecture and preserving openness to emergence.

Per Minsky \cite{minsky_marvin_l_computation_1967}, a machine -- whether it be a bio-digital system or conventional computer -- is a physical model of underlying abstract processes. But what are these underlying abstract processes that form a basis for computation? This question has been explored extensively across computer science \cite{wiedermann_what_2015, goldin_church-turing_2005,zhu_intelligent_2023,maclennan_natural_2004}, mathematics \cite{turing_computable_1937,mccarthy_basis_1963}, cognitive sciences \cite{simon_herbert_a_information_1964,newell_allen_unified_1994}, cybernetics \cite{wiener_cybernetics_1948}, and philosophy \cite{shannon_xxii_1950,copeland_what_1996,piccinini_information_2011}. There remains disagreement on the exact basis for computation, with debates extending beyond this work's scope. For example, is computation fundamentally knowledge-generating \cite{wiedermann_what_2015}, or does it merely manipulate symbols and transform information \cite{turing_computable_1937,minsky_marvin_l_computation_1967}? How do computation, cognition, and information processing relate \cite{piccinini_information_2011, goni-moreno_angel_biocomputation_2024}? 
Must we decompose computation into (hierarchical) representations at all \cite{brooks_intelligence_1991}? These debates sparked multi-hour discussions within our team, leading us to examine existing taxonomies and resources -- such as those for human computation \cite{quinn_human_2011} and AI systems \cite{yildirim_creating_2023} --  for grounding. Ultimately, we recognized that each computational framework reflects its creators' perspectives and purposes. There is no singular, ``correct'' categorization of computation -- only different lenses that reveal different possibilities.

In formulating our taxonomy, we build upon established principles from information processing theory \cite{simon_herbert_a_information_1964,goni-moreno_angel_biocomputation_2024}, abstractions from cybernetics \cite{wiener_cybernetics_1948}, and mechanistic frameworks from computer architecture \cite{turing_computable_1937,von_neumann_first_1945}. We borrow a functional modelling approach from systems design \cite{stone_development_1999,erden_review_2008,cheong_biologically_2011}, avoiding commitment to any particular computational paradigm. Recognizing that in living organisms, functions and components do not share the same one-to-one mapping or hierarchies as they often do in artificial systems, in the taxonomy we put forth here, we avoid specifying component hierarchies, instead offering a ``flat'' categorization of core computational functions. 

\subsection{Foundational Taxonomy Layers}
Our proposed taxonomy comprises eight computational layers that capture essential functions applicable across diverse computational architectures: \textbf{Input, Transduction, Evaluation/Comparison, Routing/Selection, Memory/State, Adaptation, Output,} and \textbf{Power}. Table \ref{tab:layers} describes each layer's primary functions. We characterize the functionality of these layers across two dimensions:

\textbf{Space} -- the spatial characteristics required to elicit the computational function (e.g., for organisms, (sub)cellular, organism, population, or ecosystem).

\textbf{Time} -- the temporal characteristics of the function's response, including response latency and whether effects are transient or persistent.

\begin{table}[b]
\small
  \caption{Functional layers of our computational design taxonomy for bio-digital systems}
  \label{tab:layers}
  \begin{tabularx}{\columnwidth}{lX}
    \toprule
    Layer & Function(s)\\
    \midrule
    Input & Provide electrical, chemical, physical, optical, or other signals to the system \\
    Transduction & Convert information between different representations and forms\\
    Evaluation & Determine relationships, classify information, perform comparisons\\
    Routing & Direct information flow, select paths, coordinate system components\\
    Memory & Store and enable retrieval of information and machine state\\
    Adaptation & Modify system behaviour based on experience and feedback\\
    Output & Manifest system state as electrical, chemical, physical, optical, or other signals\\
    Power & Sustain the energy requirements of computational activities\\
  \bottomrule
\end{tabularx}
\end{table}

This taxonomy extends the \textit{Living Bits} framework \cite{pataranutaporn_living_2020} by decomposing its broad concept of ``Process'' into discrete functions: Transduction, Evaluation/Comparison, Routing/Selection, Memory/State, and Adaptation. We also expand beyond the \textit{Living Bits} framework's ``Speed'' dimension to capture emergent temporal qualities more comprehensively. 

This taxonomy finds tractability in conventional computing, where it can be used to describe fundamental algorithmic operations, as well as contemporary machine learning algorithms. Notably, it also maps to diverse hardware architectures, such as Von Neumann (digital), analogue, neuromorphic, and cellular automata computers, implying a toolbox of concrete implementation pathways for each functional role. Appendix \ref{app:tradcompute} details how the taxonomy describes these operations and hardware architectures. Multiple functions may be performed by the same physical component, or multiple components may also coordinate to perform a single function; this is a notable feature that becomes especially relevant when applying the taxonomy to living organisms. 

While we concede that this taxonomy may not describe every possible computational system, we aim to demonstrate that it is still a practical starting point for analysing computational roles across hybrid architectures. We discuss in subsequent sections of this paper how this taxonomy was applied, and complicated, during our analysis of existing bio-digital systems.

\section{Analysing Existing Bio-Digital Systems}
\subsection{Methodology Overview}\label{sec:methsummary}
To investigate if and how our computational taxonomy reveals actionable insights about bio-digital systems, we applied it to analyse a scoped subset of bio-digital systems that are of particular interest to HCI researchers and practitioners. We focused on self-contained systems in which digital components physically interface with microorganisms (e.g., bacteria, protozoa, fungi, slime moulds) and DNA-based components. These biological components have practical advantages for HCI: they can be cultured rapidly (on the order of days, or faster), maintained without specialized equipment or care facilities, and integrated into diverse form factors of various scales. While plant and animal-based systems are also of interest in HCI and offer rich interaction possibilities, their development timescales and scalability place different constraints on design. Still, we believe that our taxonomy remains applicable to these systems, as we discuss in Section \ref{sec:future}. 

Our analysis was guided by the question: What computational roles do biological organisms currently play in bio-digital systems, and what patterns emerge from this distribution? To address this, we undertook a four-phase process:
\begin{itemize}
\item \textbf{Phase 1: System Collection}. We gathered a list of bio-digital systems from academic databases, review papers, printed media, and artistic portfolios.
\item \textbf{Phase 2: Screening}. We narrowed down the collection using predefined inclusion and exclusion criteria. We supplemented the collection with a secondary screening of academic and non-academic sources.
\item \textbf{Phase 3: Coding}. We coded the remaining systems using our computational design taxonomy, identifying computational roles for biological and digital components, as well as the interactions between them.
\item \textbf{Phase 4: Analysis}. We visualized the complex data with an interactive visualization platform that we created, enabling us to identify underlying patterns more readily and analyse them within our team. 
\end{itemize}

Subsequent sub-sections detail these phases. Resulting patterns are presented subsequently in Sections \ref{sec:patterns}.

\subsection{Phase 1: System Collection}
We identified relevant bio-digital systems\footnote{While we primarily use ``systems'' in the rest of this paper for consistency, most systems we analysed refer to themselves as ``artefacts,'' the more common terminology in biodesign and art.} across HCI, biodesign, and bio-art through a multi-source approach. We searched the ACM Digital Library and IEEE Xplore as primary databases for work in HCI, engineering, and biology, supplementing these with thorough reviews of included citations in key review papers in biodesign and HCI \cite{karana_living_2020, pataranutaporn_living_2020, zhou_habitabilities_2022, kim_surfacing_2023, groutars_designing_2024, ikeya_aesthetics_2025}. Additionally, we included bio-digital systems from the first author's personal collection, art and science books, online collections of bio-art exhibitions, portfolios of prominent bio-artists, and relevant review papers. This multi-source approach was essential, as many bio-digital systems emerge from artistic practice in addition to traditional academic venues. A complete list of the books and portfolios reviewed can be found in Appendix \ref{app:artefactsources} (Table \ref{tab:extra_sources}).

Our search strategy balanced comprehensiveness with feasibility. We limited our searches to 2005–2025 to capture 20 years of bio-digital system development while maintaining tractability with modern computational paradigms and biodesign tools. Only English-language, non-retracted publications were considered. Search terms were made intentionally broad to capture the diverse vocabulary used across disciplines. Table \ref{tab:search_strings} lists the complete set of search strings used for academic databases.

This collection phase took place over a 14-week period (May – August 2025). In addition to 6 review papers, 6 hardcopy books, and 6 artist portfolios, we gathered 1,500 records (ACM: 1000; IEEE: 500), of which 3 duplicates were removed.

\subsection{Phase 2: Screening}
% Need to specify why we did not use plants but we did use DNA (scalability) 
% Overall, there should be a solid reasoning for everything

We applied exclusion criteria to the titles, abstracts, and metadata of all collected literature, removing works that:
\begin{itemize}
    \item did not include a living organism or DNA (i.e. were purely bio-mimetic, bio-inspired, or bio-derived systems)
    \item did not include digital/electronic components
    \item used living organisms or DNA solely for making materials that are physically extracted before integration with a digital component (e.g., as physical non-living scaffolding of digital components) %\zb{felt like we adjusted this more yesterday? or did we do it on paper?}
    \item used digital components solely for climate control %(temperature or nutrient maintenance without computation or interaction) 
    \item used whole, multi-cellular animals (rodents, fish, etc.), plants, or organs
    \item operated solely in a clinical (therapeutic/pathology-oriented) or \textit{in vivo} context
    \item were purely theoretical or speculative pieces without functional prototypes 
    \item lacked proper descriptions of the actual organisms involved and their functions or activities
    %or descriptions of metabolic activities\ks{meaning without functional prototypes or without descriptions of metabolic activities? or meaning were without functional prototypes, or the study was just a description of metabolic activity? if the latter, make a separate bullet point} \zb{Not sure where the descriptions of metabolic activities came from... The only works that would fit under this are artworks where the organism was underspecified? Otherwise we would do a dive into the metabolic activities ourselves... }\ks{ok I'm soft-killing it :)}
\end{itemize}

This screening excluded 1,370 records, leaving 127 articles for full-text examination (83 from ACM, 44 from IEEE). A full-text examination yielded 34 articles that met all criteria. Our screening of systems cited in review papers yielded 5 additional articles. Appendix \ref{app:excluded} provides detailed examples of excluded cases.% and the reasoning behind their removal \zb{not really reasoning right}.

We applied the same inclusion/exclusion criteria to the collected bio-digital systems from the personal collection, books, and portfolio sites, supporting the process with additional documentation about the projects that was available online. From this, we identified 27 systems meeting our criteria (personal collection: 9, books: 6, artist portfolios: 12).

During preliminary coding, a second round of targeted searches were conducted to fill in underrepresented computational areas; for example, noticing that there were almost no bio-digital systems in our initial collection utilizing the biological component for memory, we opened a targeted search, looking particularly closely for DNA-based projects. We did not, in fact, uncover DNA-based bio-digital systems meeting our criteria, but we did uncover an additional biological memory example (\textit{Archean Memory Farm} \cite{inozemtseva_archaean_2022}).

\begin{figure*}[tb]
    \centering
    \includegraphics[width=\linewidth]{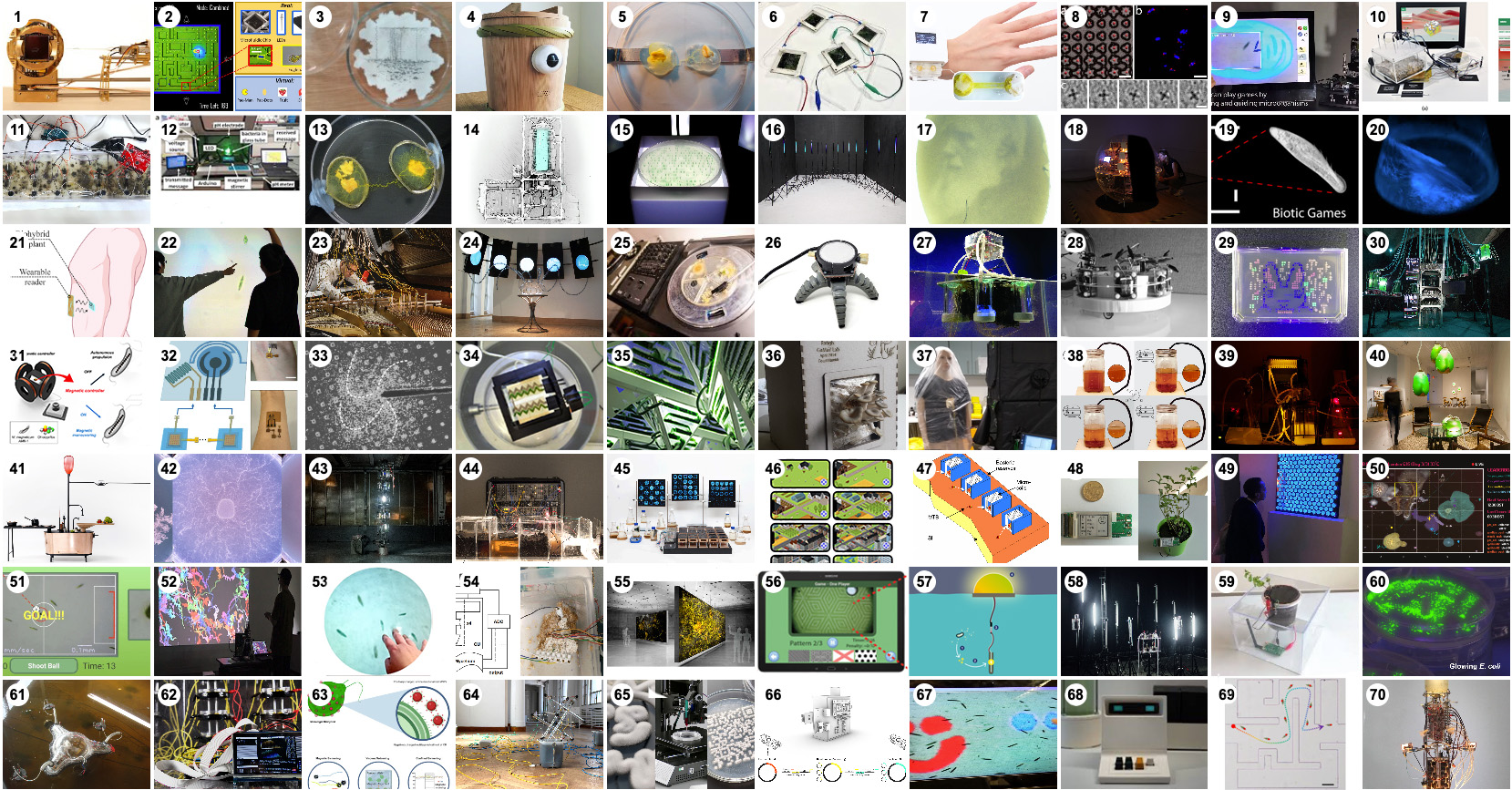}
    \caption{Images of the microorganism-based bio-digital systems selected for analysis (listed in Appendix \ref{app:included}) 1 \cite{swamp_tardigotchi_2012} 2 \cite{lam_pac-euglena_2020} 3 \cite{esparza_bio-electrolisis_2022} 4 \cite{chen_nukabot_2021} 5 \cite{whiting_slime_2014} 6 \cite{zhou_cyano-chromic_2023} 7 \cite{lu_integrating_2022} 8 \cite{zajdel_towards_2017} 9 \cite{lee_trap_2015} 10 \cite{vu_addressing_2024} 11 \cite{riggs_mold_2025} 12 \cite{grebenstein_biological_2018} 13 \cite{adamatzky_physarum-based_2016} 14 \cite{inozemtseva_archaean_2022} 15 \cite{kudla_crucible_2008} 16 \cite{interspecifics_gfp_2015} 17 \cite{tabor_programming_2007} 18 \cite{armstrong_active_2020} 19 \cite{riedel-kruse_design_2010} 20 \cite{barati_living_2021} 21 \cite{bilir_biodegradable_2024} 22 \cite{lee_microaquarium_2020} 23 \cite{miranda_composing_2018} 24 \cite{munnik_microscopic_2011} 25 \cite{team_biota_2017} 26 \cite{mishra_sensorimotor_2024} 27 \cite{henriques_caravel_2016} 28 \cite{ieropoulos_ecobot-ii_2005} 29 \cite{dobbs_bioartbot_2019} 30 \cite{sedbon_cmd_2019} 31 \cite{song_precisely_2023} 32 \cite{liu_microbial_2022} 33 \cite{pinelis_micro_2006} 34 \cite{steiner_bixels_2017} 35 \cite{ecologicstudio_urban_2015} 36 \cite{hamidi_rafigh_2014} 37 \cite{adamatzky_reactive_2021} 38 \cite{bell_bio-digital_2024} 39 \cite{c-lab_stress-o-stat_2011} 40 \cite{douenias_living_2020} 41 \cite{philips_design_microbial_2011} 42 \cite{diblasi_beauty_2023} 43 \cite{sedbon_cryptographic_2022} 44 \cite{castellanos_microbial_2017} 45 \cite{breed_algae_2024} 46 \cite{guo_slimo_2025} 47    \cite{andre_preliminary_2007} 48 \cite{rossi_let_2017} 49 \cite{c-lab_living_2013} 50 \cite{kim_new_2018} 51 \cite{riedel-kruse_design_2010} 52 \cite{lee_euglpollock_2022} 53 \cite{isitan_where_2016} 54 \cite{roberts_mining_2022} 55 \cite{puello_living_2017} 56 \cite{gerber_biographr_2016} 57 \cite{nova_innova_pond_2020} 58 \cite{interspecifics_micro-rhythms_2016} 59 \cite{rossi_long_2017} 60 \cite{cheok_empathetic_2008} 61 \cite{henriques_symbiotic_2014} 62 \cite{sedbon_c_2022} 63 \cite{akolpoglu_navigating_2025} 64 \cite{henriques_bacterbrain_2019} 65 \cite{postl_mold_2024} 66 \cite{igem_paris_bettencourt_bacterial_2022} 67 \cite{lee_tangible_2015} 68 \cite{ikeya_designing_2023} 69 \cite{wijesinghe_light-deformable_2024} 70 \cite{esparza_biosonot_2017}}
    \Description{Grid of thumbnail images for microorganism-based bio-digital systems.}
    \label{fig:allsystems}
    \vspace{14mm}
\end{figure*}

This supplementation from secondary searches added 4 systems. In total, our screening of academic databases, review papers, hardcopy books, artist portfolios, and secondary searches yielded a final dataset of \artefact bio-digital systems for coding and analysis (Figure~\ref{fig:allsystems} and Appendix~\ref{app:included}).

\subsection{Phase 3: Coding} \label{sec:coding}
The final \artefact systems were systematically coded, capturing:
\begin{itemize}
\item \textbf{Organism characteristics}, such as species, observable response, triggering mechanism (i.e., external stimuli needed to elicit response(s))
\item \textbf{Temporal characteristics} of the organism's activities of interest, such as response speed and other emergent properties (e.g., if the activity is transient or cumulative)
\item \textbf{Spatial needs}, organizational level required for detectable response (subcellular, cellular, organism, population, ecosystem)
\item \textbf{Computational role(s) of the organism} for the digital components
\item \textbf{Computational role(s) of the digital component(s)} for the organism
\end{itemize}

Deductive coding was applied for spatial dimensions and the computational role(s) of the organism, using the predetermined categories described in Section \ref{sec:taxonomy}. Inductive coding was used for organism-specific and temporal dimensions, where codes emerged during analysis.

Determining computational roles of the digital components revealed a challenge: defining roles inherently implies directionality -- that is, that the role is performed \textit{for} some agent. While ``organism'' roles are more straightforwardly interpreted to be for a digital system (and/or for a human user), ``digital'' roles required reconceptualizing what computational support means in a bio-digital system. This led to a hybrid coding approach where our predetermined taxonomy categories were supplemented with emergent codes. For example, we used ``output (connect to humans)'' in cases where digital components play the role of obviating biological activity for humans to appreciate (drawing on Zhou et al.'s taxonomy of digital tools for living organisms \cite{zhou_habitabilities_2022}) rather than directly supporting an organism's needs or assumed computational ``goals.''

\begin{table*}[t]
  \caption{Search strings used in each database}
  \label{tab:search_strings}
  \begin{tabular}{p{3.3cm}p{10cm}p{3.3cm}}
    \toprule
    Database & Search String & \# Results \\
    \midrule
    ACM DL & [[All: "microbe"] OR [All: "microbial"] OR [All: "microorganism"] OR [All: "micro-organism"] OR [All: "bacteri*"] OR [All: "cyanobacteri*"] OR [All: "slime"] OR [All: "yeast"] OR [All: "algae"] OR [All: "fungi"] OR [All: "fungal"] OR [All: "biodesign"] OR [All: "bioart"]] AND [[All: "technology"] OR [All: "electronic*"] OR [All: "biohybrid"] OR [All: "bio-hybrid"] OR [All: "interface*"] OR [All: "biodigital"] OR [All: "bio-digital"]] & 3159 (1000 taken) \\
    \addlinespace[1em]
    IEEE & ("Abstract":"microbe" OR "Abstract":"microbial" OR "Abstract":"microorganism" OR "Abstract":"micro-organism" OR "Abstract":"bacteri*" OR "Abstract":"cyanobacteri*" OR "Abstract":"slime" OR "Abstract":"yeast" OR "Abstract":"algae" OR "Abstract":"fungi" OR "Abstract":"fungal" OR "Abstract":"biodesign" OR "Abstract":"bioart") AND ("Abstract":"technology" OR "Abstract":"electronic*" OR "Abstract":"biohybrid" OR "Abstract":"bio-hybrid" OR "Abstract":"interface*" OR "Abstract":"biodigital" OR "Abstract":"bio-digital") & 1423 (500 taken)\\
  \bottomrule
  \addlinespace[1em]
\end{tabular}
\end{table*}

% TRIED
% \clearpage
% \newpage
% \pagebreak
% \cleardoublepage

The bio-digital systems were coded independently by two team members  -- one with expertise in biodesign and one in computer science -- using publications, supplementary videos, and project documentation.  Several nuances emerged during the coding that required resolution. For example, trigger mechanisms often remained implicit (e.g., a paper might describe growth without explicitly noting the addition of nutrients, the underlying ``triggers''). 
While some papers specified stimuli, others described and utilized organisms' dynamics that did not appear to have external triggers. Temporal characteristics also often required contextual interpretation, requiring inference from measurement protocols (e.g., measurement intervals), experimental timelines, or video documentation. Scale also presented challenges, as some systems utilized organisms' cellular-level activities but were implemented on a population scale; this was most notable for the biotic games. Notably, computational role boundaries were often blurred, with ``transduction'' in particular being an implicit role even when an organism appeared on first reading to be solely an input or output. We established shared criteria for handling such ambiguities. For example, organisms exhibiting autonomous behaviours, such as growth or movement, without documented external stimuli were coded with a trigger of ``none.'' 

When considering temporal characteristics, we distinguished between intrinsic and expressive timescales. While intrinsic activities, such as biochemical and metabolic activities are often rapid, the perceptible, ``expressive'' outcomes of those processes, such as changes in size or colour, often transpire more slowly. We coded organismal responses according to their expressive timescales to align with their function and expression in the context of the bio-digital system. For instance, microbial fuel cells (MFCs) generating electricity through metabolism were coded at minute-to-hour timescales to align with usable power accumulation rates and documented measurement intervals, not actual electron transfer speeds within the bacteria (which are near-instant). 

For temporal/spatial characteristics and computational roles, we applied an additive principle: for example, organisms explicitly described as ``sensors'' but performing signal transduction received both ``input'' and ``transduction'' codes, capturing a more complete computational contribution. In cases where the documentation was incomplete, particularly common among artistic works, or where the coding was ambiguous, both coders jointly searched for and reviewed available materials, discussing interpretations before reaching consensus on the final coding (example discussion in Appendix \ref{app:abm_convo}).

\begin{figure}[t]
    \centering
    \includegraphics[width=\linewidth]{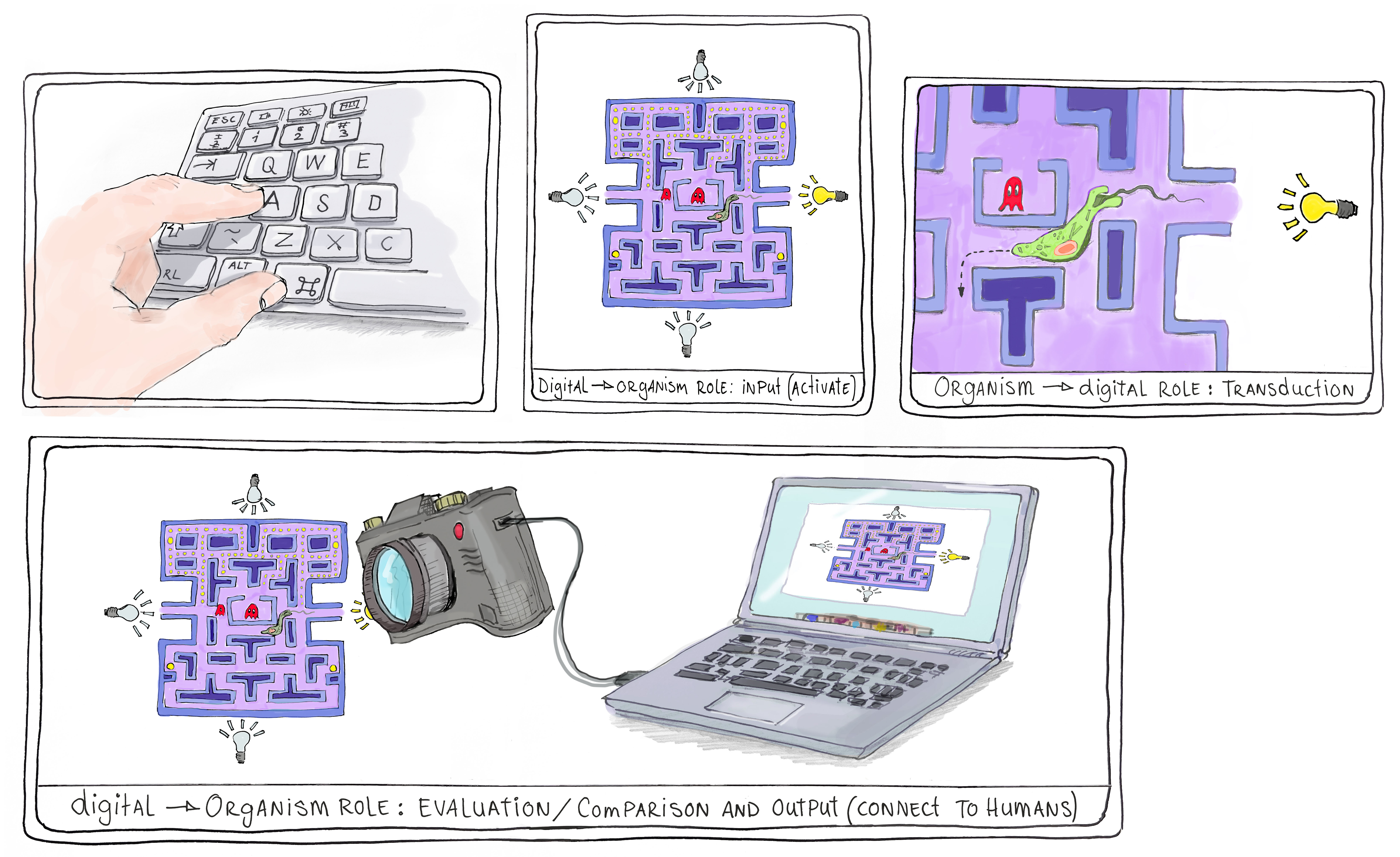}
    \caption{Interactions in the \textit{Pac-Euglena} system \cite{lam_pac-euglena_2020}. From left to right: A user presses a key; LED light is turned ON (\textit{digital $\rightarrow$ organism role: input (activate))}; the \textit{Euglena} cell moves away from the light (\textit{organism $\rightarrow$ digital role: transduction}); the movement is captured through computer vision (\textit{digital $\rightarrow$ organism role: evaluation/comparison}) and displayed back on a computer (\textit{digital $\rightarrow$ organism role: output (connect to humans})).}
    \Description{A sequence of illustrations showing interaction in the Pac-Euglena system. First, a hand presses a computer key. Second, a Pac-Man–style maze is displayed with LED lights around it, one light turned on. Third, a Euglena cell inside the maze moves away from the light. Fourth, a camera captures the Euglena’s movement in the maze. Finally, the image of the maze with the Euglena’s position is shown on a laptop screen.}
    \label{fig:pac-euglena}
\end{figure}

Here we describe Lam et. al's \textit{Pac-Euglena} \cite{lam_pac-euglena_2020} to illustrate how roles were coded (Figure \ref{fig:pac-euglena}): a human provides input through a digital interface (keyboard), which in turn stimulates the microorganism by activating LEDs (\textit{digital $\rightarrow$ organism role: input (activate)}). The organism responds, transforming the light stimulus to movement (\textit{organism $\rightarrow$ digital role: transduction}). The movement of the organism is tracked and assessed by the computer (\textit{digital $\rightarrow$ organism role: evaluation/comparison}). This activity is displayed to the humans through a projected live microscope feed (\textit{digital $\rightarrow$ organism role: output (connect to humans})).

\newpage
\subsection{Phase 4: Analysis through Interactive Visualization} 

\begin{figure*}[tb]
    \centering
    \includegraphics[width=1\linewidth]{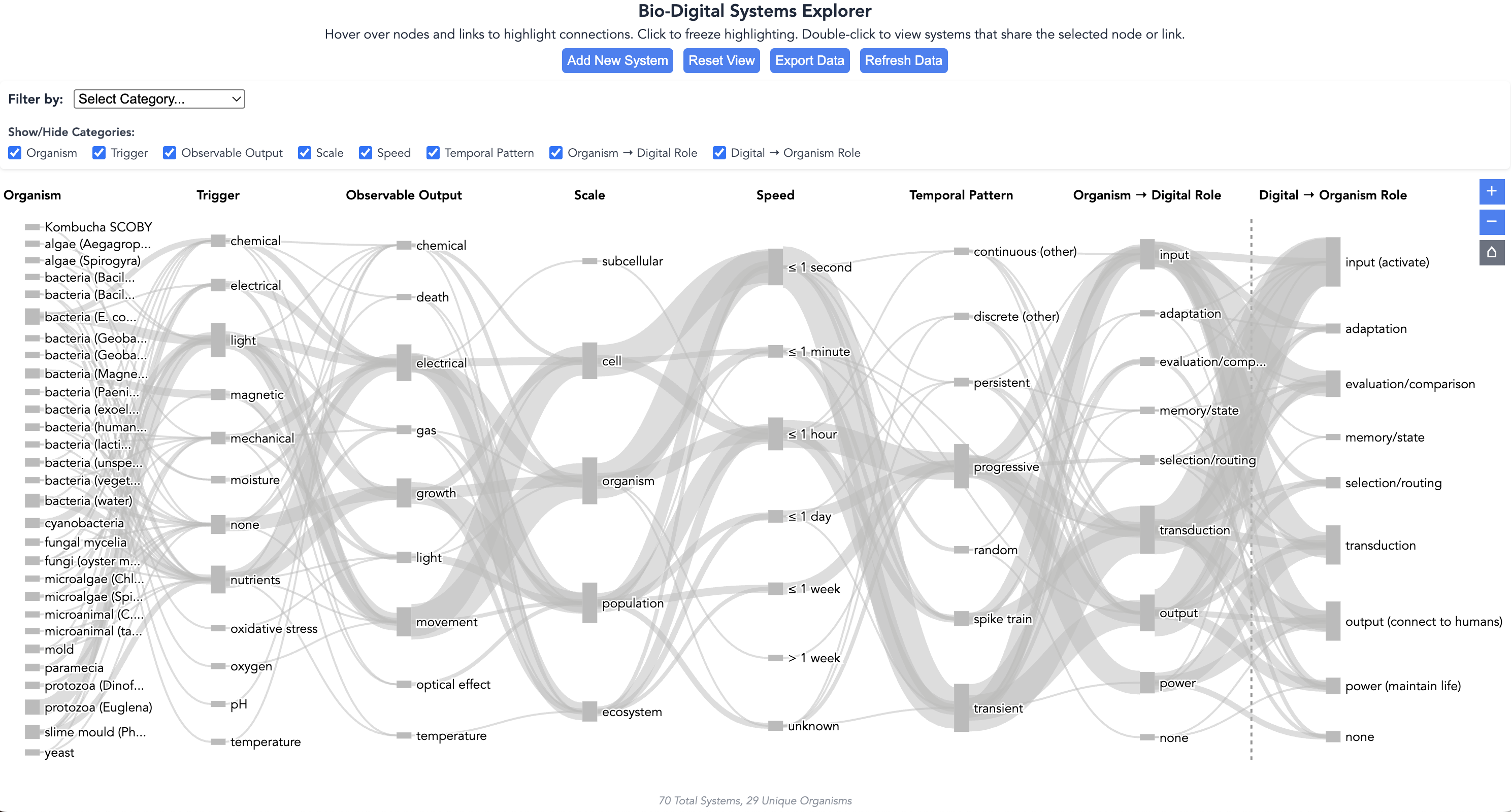}
    \caption{Our online interactive visualization platform.}
    \Description{screenshot of online visualization platform showing buttons to add new system, reset view, export data, and refresh data at top. Underneath are options to filter the data and show/hide categories. Under that is a Sankey diagram showing organism, trigger, observable output, scale, speed, temporal pattern, organism->digital role, and digital->organism role}
    \label{fig:viz}
\end{figure*}

To analyse our dataset, we developed a publicly available, interactive online visualization platform to facilitate our identification of patterns in existing bio-digital systems and identification of design opportunities\footnote{Link: https://biodigitalviz.github.io/}. The platform is implemented as a web-based application using React for interface management and D3.js for data visualization and interaction handling (Figure \ref{fig:viz}). The web app visualizes analysed bio-digital systems as a modified Sankey diagram, with columns of nodes corresponding to the dimensions of our coding, as described in Section \ref{sec:coding}.

\label{sec:viz}

Links connect nodes across different categories, with link thickness and node sizes reflecting the number of systems sharing that characteristic. This encoding makes visible the distribution of existing systems and the strength of associations between different computational roles and organism characteristics.

\subsubsection{Exploration of Bio-Digital Systems}
\begin{figure*}[tb]
    \centering
    \includegraphics[width=1\linewidth]{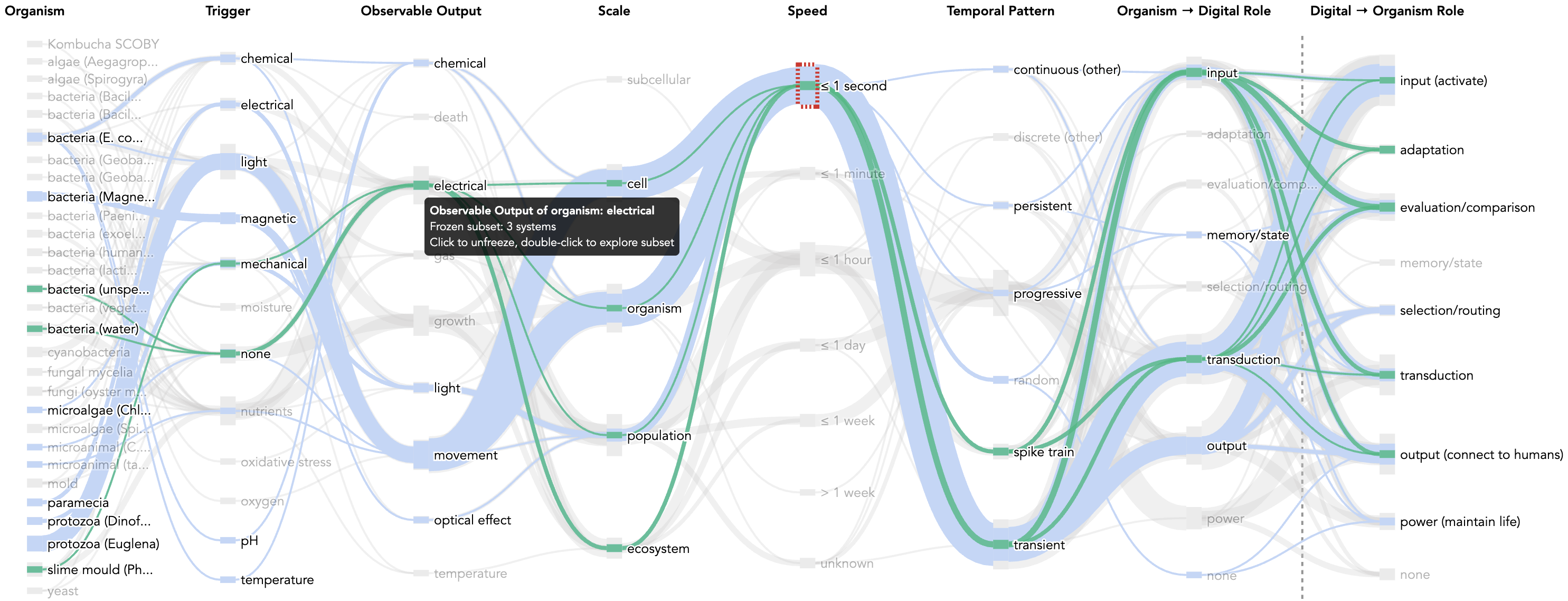}
    \caption{An example exploration using our visualization platform of the intersection between ``electrical'' observable outputs and $\leq$1 second reaction speeds. The red dashed line around the $\leq$1 second ``Speed'' node indicates that we had previously clicked on that node, highlighting connections in blue. Subsequent hovering over nodes and links, such as the ``electrical'' observable output here, highlights intersections in green.}
    \Description{screenshot of platform showing the ``frozen'' view in which the $\leq$1 second Speed node has been clicked on and encircled in a red dashed line, with faded blue lines showing all systems links that share that node. The user has hovered over the ``electrical'' observable output node, showing a green overlay highlighting that indicates 2 systems that share the intersection of $\leq$1 second and electrical output.}
    \label{fig:doublehighlighting}
\end{figure*}

To identify dominant patterns in current bio-digital systems, we identified visibly thick flows and nodes in our visualization and subsequently explored their interconnections and relationships (or lack thereof) with other nodes. We designed the platform to support multiple interactions for open-ended exploration. Hovering over nodes and links highlights complete flows through the visualization, tracing how characteristics are connected. Clicking on any node or link freezes the current highlighting, after which hovering over additional elements reveals intersections using a second colour (Figure \ref{fig:doublehighlighting}). We used this comparative highlighting to discover co-occurrences of organisms, their characteristics, and their computational roles. %Section \ref{sec:patterns} details our findings.%The visualization automatically spans links across hidden dimensions, ensuring that flows remain coherent even when intermediate columns are hidden or when the coding of an artefact is incomplete (e.g. missing information about the digital components' functional role for the organism).\ks{decided that this detail wasn't important}
\newpage

\begin{figure}[tb]
    \centering
    \includegraphics[width=0.95\linewidth]{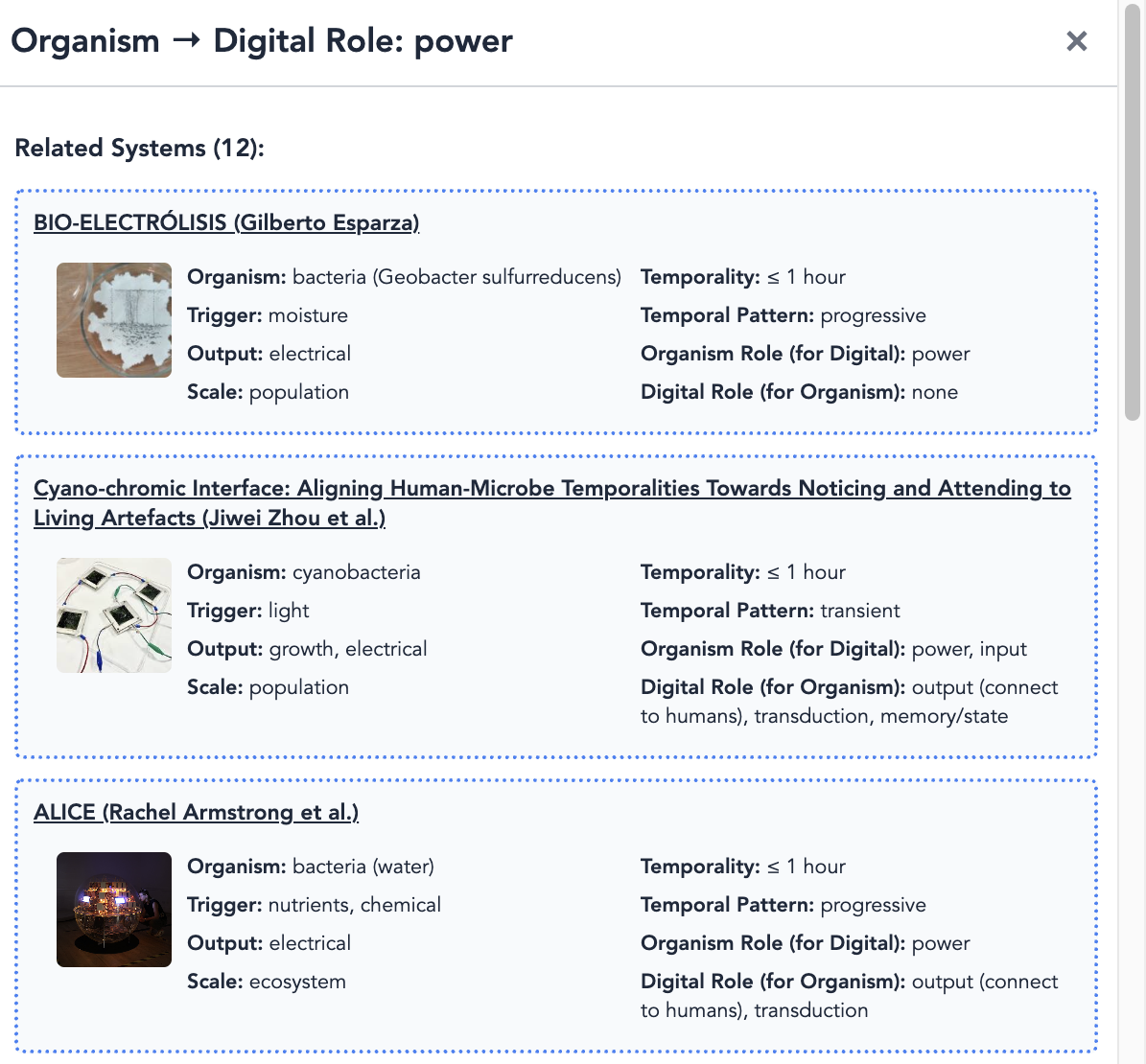}
    \caption{System Detail View for bio-digital systems in which the biological organism plays a ``power'' role for the system.}
    \Description{screenshot of detailed modal view showing a list of systems with project title, authors, organism details, triggering mechanisms, observable outputs, and taxonomic classification for the organism and for the digital components}
    \label{fig:detailview}
\end{figure}

Double-clicking a node or link opens a detailed modal displaying the systems that share that node or link (Figure \ref{fig:detailview}). Each entry details codes assigned to that system, along with a thumbnail image and an external link to its paper or webpage. When viewing intersections between frozen and hovered selections, the detail view shows only systems matching both criteria.

We also inspected the visualization for sparse connections and small nodes that indicated underutilized pathways or gaps in the dataset. We searched for gaps in both the default view and within subsets, using the previously described frozen highlighting function. Our visualization platform supports this exploration by allowing columns in to be filtered or rearranged (by dragging and dropping).

\subsubsection{Community-Driven Curation}
We have populated this visualization with the \artefact microorganism-based bio-digital systems that our team collected and coded. However, this is merely the start of an organic, community-driven repository of bio-digital systems to which other researchers can also contribute. The platform includes a form through which anyone can contribute and classify new bio-digital systems. Submissions are instantly added to an Airtable database. Upon submission of the form, new systems are also immediately integrated into the visualization. In the short term, we plan to manually monitor submissions for consistency.
% \hfill \break
\section{Results: Patterns in Current Bio-Digital Systems} \label{sec:patterns} 
%\ale{I am missing some information here (or perhaps just reminders) about what process this analysis followed. Something is described in 4.1 (especially the descriptive and Prescriptive questions), but there is no follow up}\ks{added refs/reminders...though now it's a little strange that the "results" is answering Q1, whereas Q2 is left to the first chunk of the "discussion" (this was the reason why originally it was its own section...)}
%\ek{can we remove the Q2 from the previous section?}\ks{done}

We now describe the results of our computational taxonomy-guided analysis of existing bio-digital systems. Our identification and subsequent explorations of dense and sparse areas of our visualized data reveal significant clustering patterns in how current systems leverage organisms, biological and digital computation, temporal qualities, and spatial qualities.

\subsection{Organism Features}
Biological outputs used in current microorganism-based bio-digital systems cluster around three modalities: electrical signals (36\%, 25 systems), movement (27\%, 19 systems), and growth (27\%, 19 systems). In contrast, chemical outputs (3 systems, e.g, \textit{Biodegradable Implant} \cite{bilir_biodegradable_2024}), temperature (1 system, \textit{Nukabot} \cite{chen_nukabot_2021}), and non-light-generating optical effects (2 systems, e.g., \textit{Living Mirror} \cite{c-lab_living_2013}) occur only sparingly. This clustering reflects the type of organisms commonly used in designs: nearly all movement-based systems utilize \textit{Euglena} (e.g., \textit{EuglPollock} \cite{lee_euglpollock_2022}) or paramecia (e.g., \textit{Where Species Meet} \cite{isitan_where_2016}) for their predictable phototaxis, and electrical-based systems predominantly employ electrogenic bacteria as MFCs (e.g., \textit{ALICE} \cite{armstrong_active_2020}). Systems consistently use organisms for single, well-characterized behaviours, with limited consideration of their broader, multi-modal capabilities.

Trigger mechanisms exhibit similar clustering. Light (33\%, 23 systems) and nutrients (26\%, 18 systems) are the most common stimuli, with systems leveraging well-understood biological responses like phototaxis and chemotaxis and simple digital control (e.g., LEDs for light stimulation (e.g., \textit{Pac-Euglena} \cite{lam_pac-euglena_2020}) and a syringe for nutrient delivery (e.g., \textit{Tardigotchi} \cite{easterly_tardigotchi_2010})). Despite offering comparable precision, alternative triggers such as magnetic fields (5 systems, e.g., \textit{Living Mirror} \cite{c-lab_living_2013}), chemical signals (6 systems, e.g., \textit{Reactive Fungal Wearable} \cite{adamatzky_reactive_2021}), and pH changes (1 system, \textit{Biodegradable Implant} \cite{bilir_biodegradable_2024}) remain underutilized. Notably, 11 bio-digital systems (16\%) employ no specific external trigger, relying instead on organisms' ambient responses or intrinsic metabolic variations (e.g., \textit{Microscopic Opera} \cite{munnik_microscopic_2011}).

\subsection{Computational Role Asymmetries}

\begin{figure*}[t]
    \centering
    \includegraphics[width=0.8\linewidth]{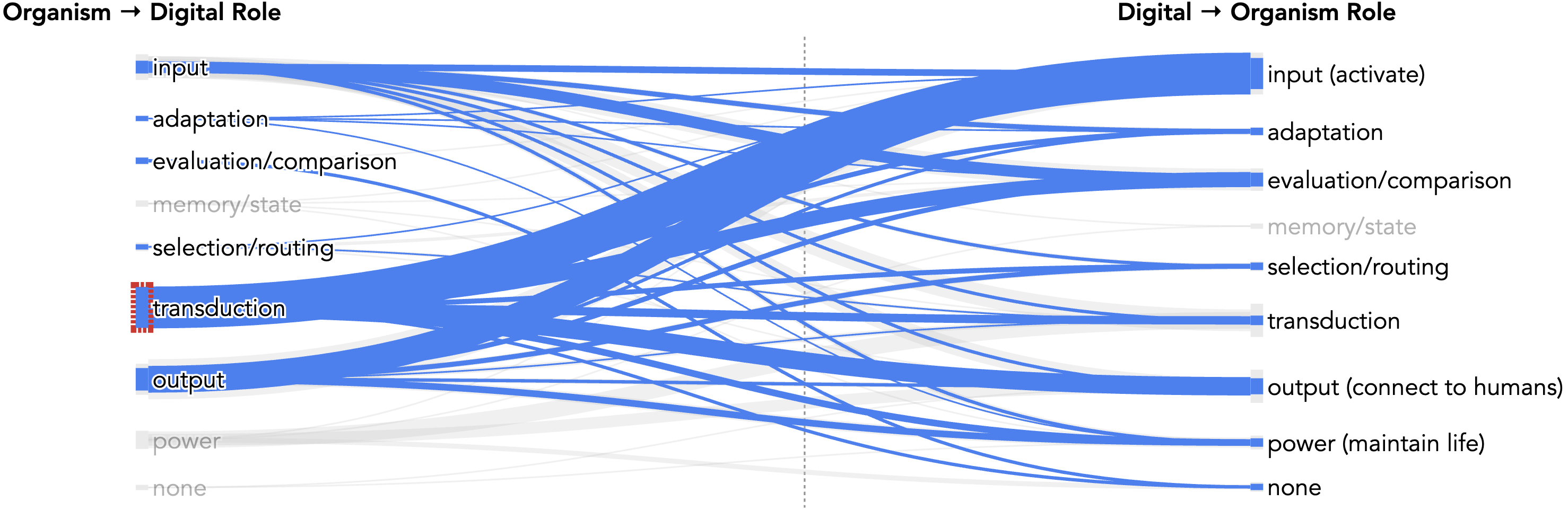}
    \caption{In 49\% of the analysed bio-digital systems, the organism serves a transduction role (encircled in red). In 50\% of the systems, the digital components serve as the input to activate the organism's response.}
    \Description{Screenshot from our Sankey visualization showing columns for organism->digital roles and digital->organism roles. Transduction in the former is the most common category, with input (activate) in the latter being the most common category there.}
    \label{fig:transformation}
\end{figure*}

Our analysis reveals fundamental asymmetries in biological and digital computational partnerships. Organisms mostly serve a transduction role (49\%, 34 systems), converting stimuli through metabolic or behavioural processes (Figure \ref{fig:transformation}). This transduction role almost invariably co-occurs with the roles of output (36\%, 25 systems) -- with organisms producing observable movement, growth, or bioluminescence -- or input (29\%, 20 systems) -- with organisms serving as sensors that produce digitally detectable responses. Meanwhile, 13 systems (19\%) use organisms for power generation as MFCs. As we expected, advanced computational roles for organisms are nearly absent: evaluation/comparison appears in 3 systems (e.g., \textit{Mining Logic Circuits in Fungi} \cite{roberts_mining_2022}), selection/routing in 4 (e.g., \textit{Algae Relay} \cite{ikeya_aesthetics_2025}), memory/state in 2 (\textit{Archaean Memory Farm} \cite{inozemtseva_archaean_2022} and \textit{Cryptographic Beings} \cite{sedbon_cryptographic_2022}), and adaptation in only 1 (\textit{Beauty} \cite{diblasi_beauty_2023}).

Digital components show inverse patterns in terms of the computational roles that they play for the organism. They serve as input providers (``activators'') to the biological system in 35 systems (50\%), providing controlled stimuli such as light, sound, or electrical signals. In 9 systems (13\%), digital components merely play a power role for the organism (e.g., \textit{Living Things} \cite{douenias_living_2020}); 2 of these serve as simple life support systems, with power being the \textit{only} role they play for the organism. They serve as output in 27 systems (39\%), translating biological output to perceivable forms for human interpretation and interaction (e.g., as a display to obviate organismal movement (e.g., \textit{Trap it!} \cite{lee_trap_2015}, \textit{BioGraphr} \cite{gerber_biographr_2016}). %\zb{maybe something to note (not sure if interesting) but output is always paired with something else}.
Transduction occurs in 27 cases (39\%) and usually co-occurs with other roles. Notably, despite capacity to perform any and all of the computational functions we define in our taxonomy, digital components also play fairly primitive computational roles to support biological computation in bio-digital systems, with only 4 systems providing adaptation roles (e.g., \textit{Composing with Biomemristors} \cite{miranda_composing_2018}, \textit{CMD} \cite{sedbon_cmd_2019}) and only 1 system providing memory/state support (\textit{Cyano-Chromic Interface} \cite{zhou_cyano-chromic_2023}). 

\begin{figure*}[tb]
    \centering
    \includegraphics[width=0.85\linewidth]{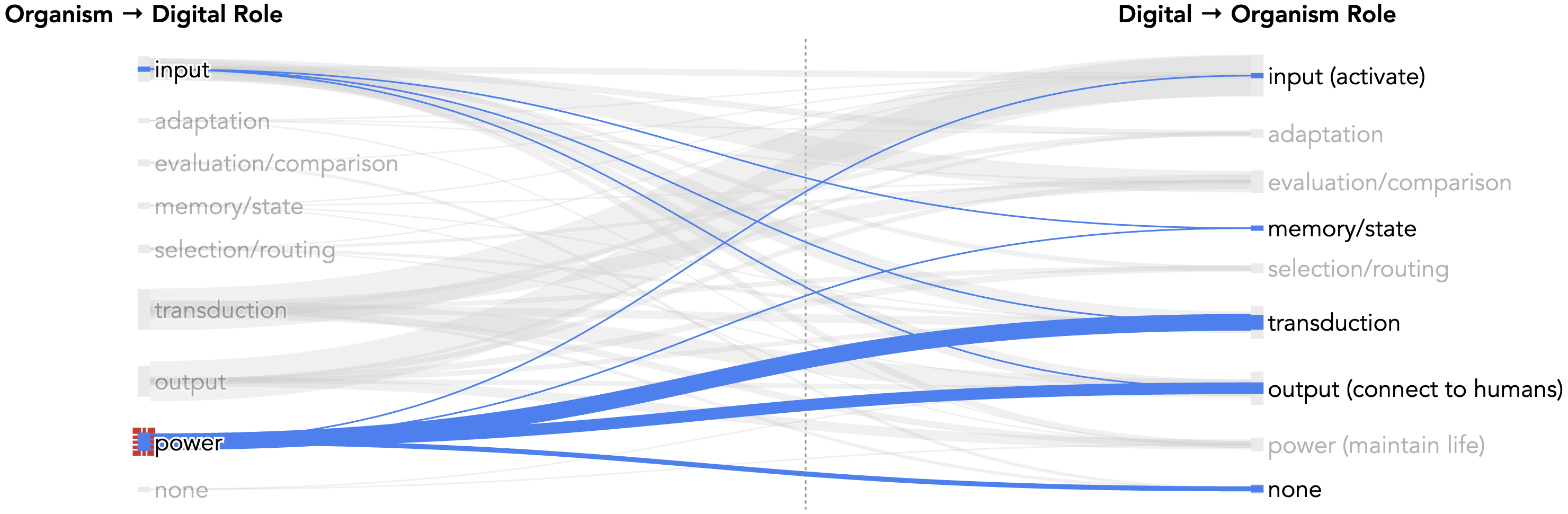}
    \caption{In systems where the organism plays a power-generating role (encircled in red), it rarely plays other computational roles (in only 1 system \cite{zhou_cyano-chromic_2023}, it plays a second role, which is input).}
    \Description{Screenshot from our Sankey visualization showing columns for organism->digital roles and digital->organism roles. "Power" as an organism role is highlighted. The only other co-occurring role shown is input.}
    \label{fig:power}
\end{figure*}

This distribution reveals that beyond simple input or output, current bio-digital systems position organisms as merely transducers, converting digital inputs to biological outputs or vice versa. In systems where the organisms generate power, the digital components rarely leverage the organisms' other capabilities, treating them as single-function components rather than multi-faceted computational agents (Figure \ref{fig:power}). The near-absence of biological memory (2 systems) and adaptation (1 system) is particularly striking given the fact that cellular memory and evolution are fundamental biological processes. In general, our results reveal that current bio-digital systems neither fully exploit the inherent strengths of each component nor provide a platform for genuinely reciprocal interactions between them.

%In systems where the organisms generate power, the digital components rarely leverage the organisms' other capabilities, treating them as single-function components rather than multi-faceted computational agents.

%\ks{to-do--these sentences are a bit misplaced (is this actually notable?)}Notably, in nine artefacts, the role of the digital is \textit{power (maintain life)} whereas in five systems the digital role is \textit{none}. Conversely, for the role of organism for digital, only one systems has the role \textit{none}.

% This asymmetric relationship reveals that current bio-digital artefacts leverage each component's inherent (not inherent I think because I would say that is not really true) strengths. Digital precision for input generation and biological complexity for transformation and output. Rather than creating truly reciprocal interactions.

\subsection{Temporal Qualities of Bio-digital Systems}

\subsubsection{Response Time/Speed}
\begin{figure*}[ht]
    \centering
    \includegraphics[width=0.79\linewidth]{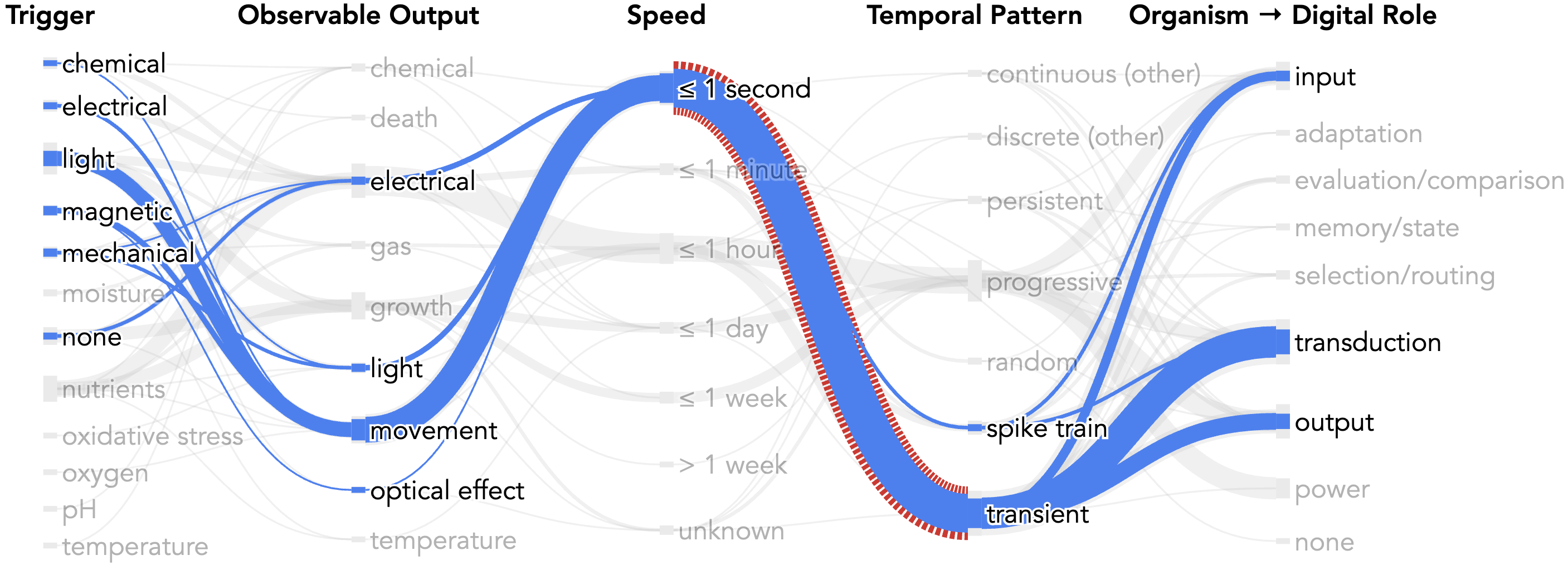}
    
    \caption{36\% of systems analysed use organisms' ``fast'' ($\leq$1 second) responses. These are almost exclusively transient responses (the intersection between $\leq$1 second and transient responses is encircled in red) that co-occur most frequently with movement as the modality of response.}
    \Description{Screenshot from our Sankey visualization showing columns for trigger, observable output, speed, and temporal pattern, and organism->digital role with the link between speed =< 1 sec and temporal pattern=transient highlighted}
    \label{fig:temporal}
\end{figure*}

The speed of utilized biological responses inversely correlates with computational sophistication. ``Fast'' organisms with expressive responses occurring in less than 1 second (e.g., \textit{GFP Screen} \cite{interspecifics_gfp_2015}, \textit{Living Mirror} \cite{c-lab_living_2013}) (36\%, 25 systems) are often simply used as outputs to the system (36\% of fast systems), alongside a transduction role (72\% of fast systems) (Figure \ref{fig:temporal}). These systems are mostly biotic games (e.g.,  \textit{Pac-Euglena} \cite{lam_pac-euglena_2020}) that utilize organisms' transient movement (dinoflagellate-based systems, which output light, are exceptions) arising in response to light or electrical triggers. Minute-to-hour timescales feature bacterial MFCs with progressive (i.e., without return to a baseline state) charge accumulation. Still, organisms at this timescale are limited primarily to transduction and power roles.

At longer timescales >1 hour, growth becomes the dominant mode of output, and more sophisticated computational roles begin to emerge. \textit{Nukabot} \cite{chen_nukabot_2021} leverages three bacterial species' fermentation processes on an ecosystem scale, while \textit{Beauty} \cite{diblasi_beauty_2023} bestows more complex computational roles onto organisms, such as adaptation. Still, advanced computational capacities such as memory, state management, and adaptation remain rare across all timescales.

\subsubsection{Response Patterns}
49\% of systems analysed (34 systems) utilize transient effects -- ones returning to baseline in the absence of a stimulus. 44\% (31 systems), utilize progressive effects -- ones accumulating over time. Systems exhibiting non-transient or non-progressive patterns -- such as persistent (i.e., an effect that is sustained even after the removal of a stimulus) and random patterns (e.g., noise) -- are rare, collectively representing only 7\% (5 systems) of the dataset.

Some patterns always co-occur. For example, transient patterns correlate with fast reaction times (64\% of transient patterns) and particular organism output modalities: movement (e.g., \textit{Pac-Euglena} \cite{lam_pac-euglena_2020}), followed by electrical (e.g., \textit{Micro-Rhythms} \cite{interspecifics_micro-rhythms_2016}) and light (e.g., \textit{Algae Alight} \cite{breed_algae_2024}). Progressive patterns correlate with long reaction times (with the exception of \textit{Archaean Memory Farm} \cite{inozemtseva_archaean_2022}) and growth (e.g., \textit{The Crucible} \cite{kudla_crucible_2008}) or electrical outputs (e.g., \textit{ALICE} \cite{armstrong_active_2020}). Additionally, spike train patterns are inherently also transient and electrical (e.g., \textit{BioSoNot} \cite{esparza_biosonot_2017}).

\subsection{Spatial Qualities of Bio-Digital Systems }
Most bio-digital systems operate at single-species scales (83\%, 58 systems), utilizing the activities of single organisms or of a population of a single species. The biological component is often also contained in a small physical size, for instance as a bacterial colony in a Petri dish (e.g., \textit{Beauty} \cite{diblasi_beauty_2023}) or within a microfluidic device (e.g., \textit{Pac-Euglena} \cite{lam_pac-euglena_2020}). This physical constraint was so universal in our dataset that we excluded physical size from our coding scheme -- virtually all systems operated at bench-top scale or smaller, with rare exceptions using projections to visually amplify microscopic organisms (e.g., \textit{MicroAquarium} \cite{lee_microaquarium_2020}) or artistic installations. 

%making the interaction with the digital components tightly coupled and spatially constrained. 

The 12 systems that do operate at the ecosystem level mostly rely on naturally occurring microbial ecosystems, such as those found in water or in the human body. However, they do not explicitly leverage connections or interactions between these organisms. Even within these ecosystem-scale systems, organisms are typically confined within small spatial footprints (e.g., \textit{BioSoNot} \cite{esparza_biosonot_2017}), functioning in closed ecosystems. \textit{POND} \cite{nova_innova_pond_2020} is one system that operates in an open ecosystem, utilizing the activities of organisms in the surrounding environment, but its digital (and mechanical) components are still physically confined. %Other systems such as \textit{Carvel} \cite{henriques_caravel_2016} have been designed with the potential to operate in an open ecosystem but are not yet employed in such context.  
\section{Discussion} 
%\ks{from Elvin: when we share an example in this section, we could also show the envisioned artefact with all the taxonomy levels overlaid onto it}
\label{sec:opportunities}
Our taxonomy-guided analysis revealed computational blind spots that guide us towards targeted explorations for future bio-digital systems. We subsequently highlight a few of these gaps and describe design pathways for more sophisticated and synergistic bio-digital systems. We further discuss the contributions such systems could make to HCI and to the broader discourse on regeneration. 

\begin{figure*}[t]
    \centering
    \includegraphics[width=15cm]{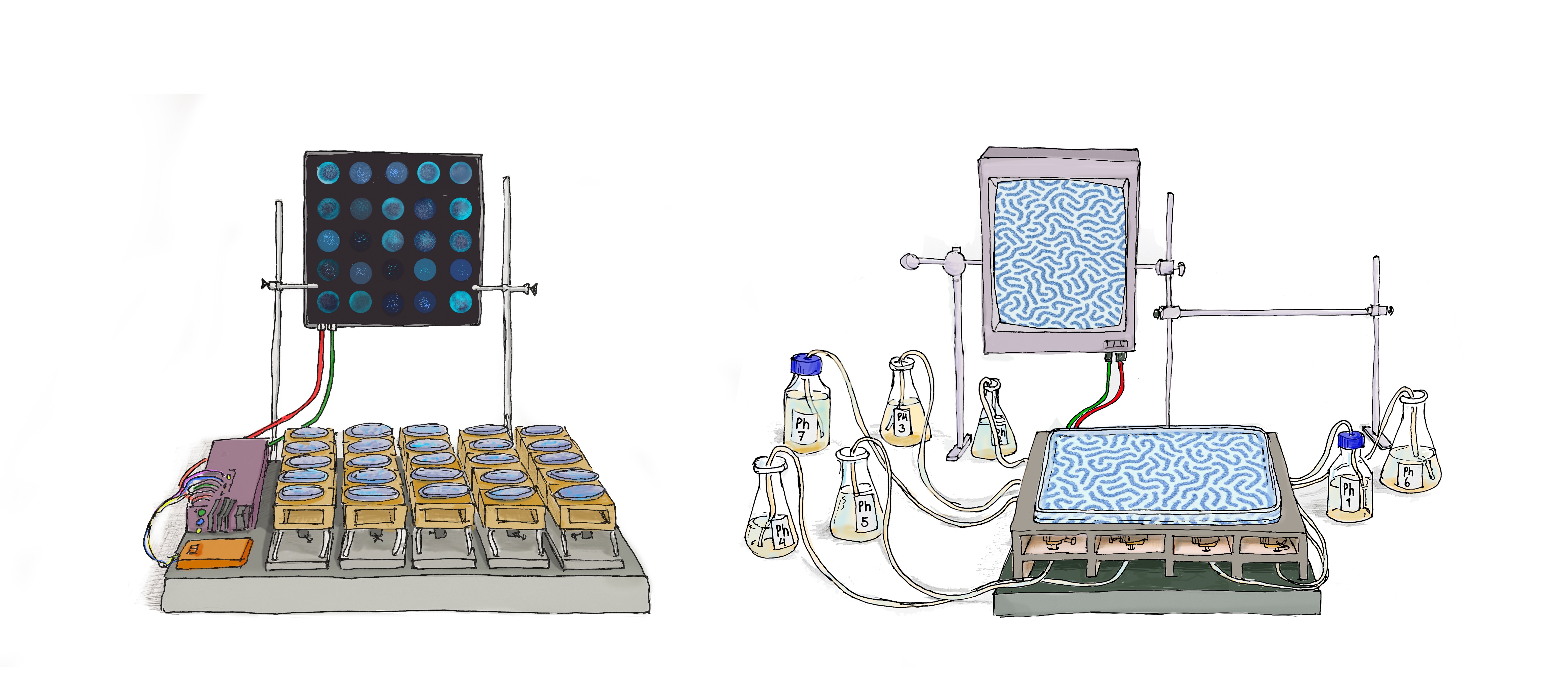}
    \caption{Illustration of how the \textit{Algae Alight} system \cite{breed_algae_2024} might be transformed by incorporating multi-modal triggers. On the left, the original system is shown, visualizing an array of petri dishes holding dinoflagellates excited by mechanical shakers. On the right, an extended concept incorporates pH-based stimulation of dinoflagellates. Different pH solutions can be delivered to precise points in the dinoflagellate tank. A display visualizes the organisms’ responses to these inputs, indicating that fine-grained control can be achieved through targeted pH modulation.}
    \Description{This figure presents an extended concept for the Algae Alight system, showing how pH-based stimulation can be incorporated into existing bio-digital systems. On the left, the original set-up is depicted, while the right side illustrates a modified version that enables modulation of bioluminescent responses through targeted pH inputs. Around the system, bottles containing solutions of varying pH serve as inputs. These chemical additions locally alter the environment, eliciting bioluminescent responses that are visualized on a display positioned adjacent to the system. By linking the organisms’ reactions to controlled pH changes, the concept demonstrates how fine-grained, spatially differentiated interaction can be achieved.}
    \label{fig:dino_pH}
\end{figure*} 

\subsection{Untapped Potentials of Bio-Digital Systems}

\subsubsection{Multi-Modal Organism Capabilities}
Our results show that microorganisms in bio-digital systems are often reduced to a simple function, with a single trigger producing a single output, despite their inherently complex, multi-modal processes. Triggering mechanism cluster around nutrients and light, while alternative modalities -- such as temperature or pH -- remain underexplored. Using a menu of diverse triggers in concert could enable systems that can activate both large-area, population-wide responses and small-scale, precise responses.
%\ks{what are context-specific organism responses?} \zb{indeed a bit vague and empty, what I meant to say is that they are tailored to or indicative of environmental conditions/situations instead of just an on/off reaction orrr predetermined output.. such as shine light --> organism moves} \zb{changed it but maybe the example after is now less fitting/strong}
%\zb{we could also add something like: using stimuli that mirror the actual conditions organisms adapt to in nature (e.g. oxygen gradients, temperature changes), but then what is the advantage of this?}\ks{ok got it. I moved that part to the end of the next paragraph...you also gave me the idea that one potential advantage is that some triggers allow for "fine-grained" control, while others you kind of have to expose the whole population} \zb{nice, really good addition}

Dinoflagellates, a micro-algae commonly used in bioluminescent systems, exemplify this untapped potential. While current dinoflagellate-based bio-digital systems rely on mechanical agitation to trigger light emission (e.g., \textit{Algae Alight} \cite{breed_algae_2024}, \textit{Living Light Interface} \cite{barati_living_2021}), dinoflagellates also produce bioluminescence in response to pH modulation, pressure variations, electrical stimulation, and exposure to specific ions \cite{hamman_mechanical_1972}. These capabilities could be harnessed for sensing in conditions where electronic pH meters and visual indicators interfere with processes of interest or are cumbersome to protect against the environment. A pH-triggered dinoflagellate system could, for instance, be coupled with various electronic or biological activities (e.g., pH manipulation, metabolic activity of other organisms, or sensor-controlled acid/base release) to enable multi-modal systems with localized responses that more fully harness the organisms' potentials (Figure \ref{fig:dino_pH}). Organisms could be triggered to convey fine-grained responses or spatio-temporal context, such as growth patterns that reveal the evolution of local thermal dynamics over time or biochemical outputs that map chemical compositions (e.g., pollutant concentrations).

%, moving beyond mere passive response to motion created by mechanical stimuli.\ks{I commented this out because it was still not clear. does "passive response" = light? why is it significant that we are moving from getting them to produce light to getting them to move?}

% Our results furthermore show that designers of bio-digital systems often reduce microorganisms to a single function, despite their inherently multi-modal capabilities. 

We similarly observe that current bio-digital systems utilize only one single output modality of the organisms they incorporate. They particularly overlook the rich design space offered by biochemical signalling -- with only three systems making use of chemical outputs (\textit{Nukabot} \cite{chen_nukabot_2021}, \textit{Biological Optical-to-Chemical Signal Conversion Interface} \cite{grebenstein_biological_2018}, \textit{Biodegradable Implant} \cite{bilir_biodegradable_2024}).
%\zb{to do: add info below from discussion pt 2. into this text}
%Reducing organisms to a single function not only represents a missed opportunity for design innovation but also risks using living resources in reductive and extractive ways, undermining regenerative ambitions. Organisms offer unique computational functions -- from self-repair and growth to cyclical material and energy exchanges -- that could inspire or directly enable digital systems to operate in more regenerative ways for the benefit of larger ecosystems. \zb{hmmm not sure because the example we give after it is not directly linked to this}\ks{I agree this is strange...I think this section stands without it actually (I think we mention this kind of thing elsewhere?), so I'm soft-killing...}

To unravel new possibilities for combining and broadening modalities, we took a closer look at one of the commonly used organisms in our collection: \textit{Physarum polycephalum} (slime mould). Existing \textit{Physarum}-based systems utilize its motility and electrical oscillations, which are linked to cytoplasmic streaming. However, \textit{Physarum} also exhibits chromatic shifts in response to pH changes. While normally yellow, it is deep red-orange in acidic conditions (pH$\approx$1) and bright yellow-green in alkaline environments (pH>8) \cite{adamatzky_physarum_2010, seifriz_slime_1935}. Considering these capacities together, we began to imagine how a system might take a dual-parameter approach, for example, by harnessing the organism’s electrical oscillations as indicators of internal computational states, while using its pH-responsive colour shift as an immediately perceptible environmental feedback channel. Such a system could function as a chemical sensor that embodies both detection and communication functions within a single bio-digital system. More broadly, utilizing multiple stimuli can enable a broader range of spatial and temporal control; for instance, mechanical stimulation typically tends to operate at centimetre-scale resolution, while pH-based cues can be tuned to millimetre-scale precision.

% \new{The broadening of organismal modalities opens possibilities for multi-layered forms of interaction that reflect established HCI work on multimodal interfaces, meeting the need for richer and more expressive feedback mechanisms in living systems.}

% \zb{Currently feels a bit shallow. Should it be more of a concrete example? or we could go the context-aware route (interaction models built around context-aware...}\ks{this is too vague and not the right angle (also very genAI)...Multimodal interfaces mean engaging [humans'] taste, smell, etc., which I think isn't as relevant. 

% Better to say something along the lines of utilizing multiple stimuli -> enabling a wider range of temporal and spatial control (e.g. mechanical stimulation's resolution is probably on the order of cm; pH can be probably on the order of mm?)}

% Importantly, reducing organisms to a single function not only represents a missed opportunity for design innovation but also risks using living resources in reductive and extractive ways that undermine regenerative ambitions. Organisms offer unique computational functions -- from self-repair and growth to cyclical material and energy exchanges -- that could inspire or directly enable digital systems to operate in more regenerative ways for the benefit of larger ecosystems. To that end, our study also revealed significant untapped potential for bio-digital systems to contribute to regeneration through designing across scales.\ks{commented out. I think elements of this are now sprinkled into different sections...}

\subsubsection{Biological Memory x Digital Routing}
\begin{figure}[h]
    \centering
    \includegraphics[width=8.5cm]{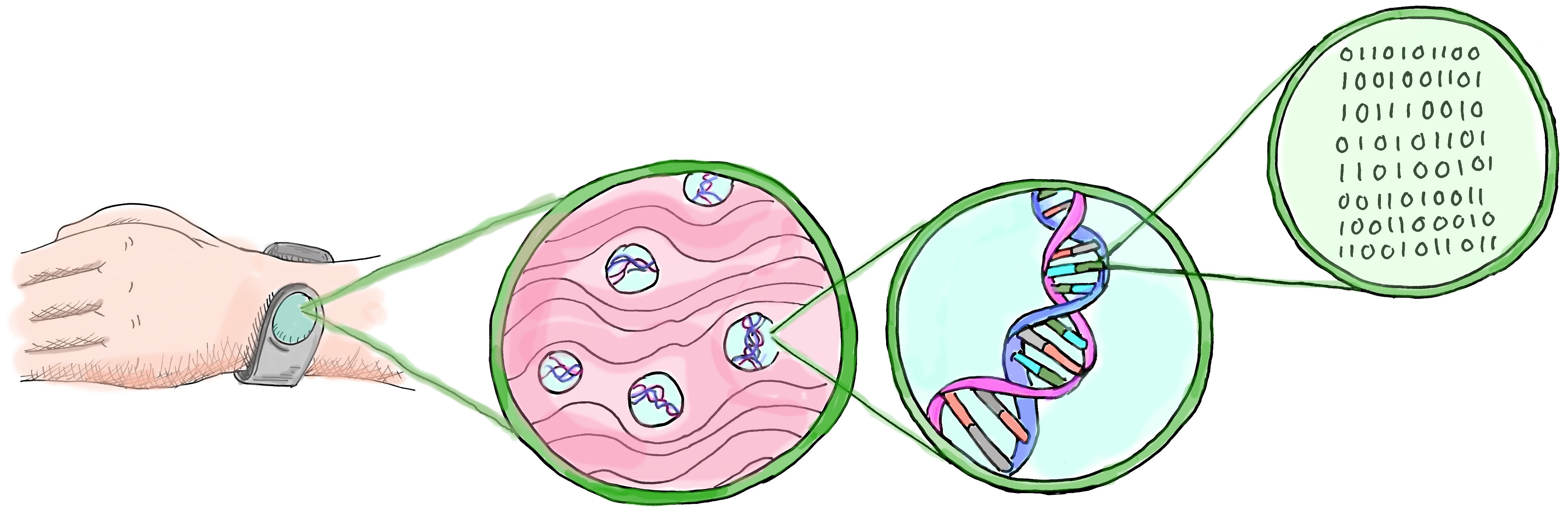}
    \caption{Illustrative concept of a bio-digital system combining DNA memory with digital routing. A composite material interfaces digital interconnects with embedded DNA-containing particles. The molecular DNA strand densely encodes information. The digital interconnects enable access (reading and writing) to the information stored in the DNA.}
    \Description{Illustration of a conceptual bio-digital system. A wrist worn device displaying zoomed-in views: first, a material embedding particles with enclosed DNA; next a close-up of the DNA helix; finally, binary code representing digital information encoded in DNA.}
    \label{fig:dna-memory}
\end{figure}

We did not identify any systems combining biological memory/state with digital selection/routing. Pursuing this combination could enable distributed biological memory, where information is stored or encoded in living systems and selectively accessed or directed through digital controls. \textit{Archaean Memory Farm}, which proposes using the accumulation of dead magnetotactic bacteria as memory \cite{inozemtseva_archaean_2022}, and \textit{Cryptographic Beings}, which uses the floating and sinking of \textit{Aegagropila linnaei} (algae balls) as bi-stable ``bits'' of memory \cite{sedbon_cryptographic_2022}, could be advanced by integrating digital systems to coordinate and enhance information access. Additionally, DNA-based storage has already been demonstrated at scale, with entire books, images, and even operating systems successfully encoded into DNA strands \cite{ceze_molecular_2019}. While somewhat unstable at high temperatures, DNA offers unique advantages: it is shape-independent, self-replicating, and orders of magnitude more physically dense than conventional storage devices \cite{koch_dna--things_2020}. We did not find any DNA-based bio-digital systems meeting our selection criteria (i.e., ones that that physically interface DNA with electronics), but researchers \cite{koch_dna--things_2020}, and even companies \cite{biomemory_building_2021}, have developed read-out techniques and digital sequencing machines that future work could integrate into self-contained bio-digital systems. Pursuing integrated digital routing could enable more synergistic bio-digital systems that not only take advantage of but also forefront the vast capabilities of DNA storage for HCI applications (Figure \ref{fig:dna-memory}). 

For interaction design, biological memory offers many potential advantages. For example, DNA can be embedded within flexible, miniaturized, or unconventional form factors in environments where conventional rigid memory systems are impractical or undesirable \cite{koch_dna--things_2020}.

%\new{For interaction design, biological memory offers (quick) adaptive advantages, enabling systems to respond to changing conditions and guide behaviour based on past experiences, (without requiring direct learning??) encoded in living substrates.}

% \zb{Integrating some of this below...  Biological memory provides a advantage by enabling learning, quick adaptation to threats, and efficient recall of beneficial or dangerous past experiences. On top of that biological memory, unlike simple stimulus-response circuits, is integrated into an organism’s wider cognitive economy. Biological memory does not seem to require learning via direct experience}\ks{this is too abstract imo...i.e. what does it even mean to have an adaptive memory? maybe for now stick to the advantages re. form factor}

% Humans, after all, routinely read and act upon instructions, maps and manuals written by others, drawing on information acquired through their experiences, not our own. Although such externalised sources of information are typically declarative in structure – designed to represent facts explicitly – we often act upon them automatically, without needing to consciously recall or reflect on the information they convey. In this way, they guide behaviour in ways that functionally resemble non-declarative memory.

% \ale{This is perhaps a limitation of your papers sample? There are companies (https:\/\/www.biomemory.com) that offer the service. And, in general, interfacing DNA with electronics means having a sequencing machine (for reading) and then do something with it. } 

\subsubsection{Biological Evaluation x Digital Adaptation}
The systems that we identified that utilize organisms for evaluation/comparison (e.g., slime mould in \cite{whiting_slime_2014,roberts_mining_2022}) do not leverage digital components' full range of computational abilities to support the integrated organism. In pursuit of this gap, we identified one opportunity: physical reservoir computing, an emerging computational approach that leverages substrates with complex, non-linear dynamics and temporal memory -- traits common in many biological systems \cite{nakajima_physical_2020}. 
Physical reservoir computing could harness bacterial collective decision-making or slime mould path optimization as evaluation/comparison, for example by processing spatio-temporal inputs, such as gestures. Digital components could provide adaptation, learning optimal stimulation protocols based on real-time biological responses. Recent works within this context demonstrate potential. Scientists have achieved impressive accuracy and significant resource savings on benchmark classification tasks by modelling reservoir networks after mycelium growth \cite{tompris_mycelium_2025}, solving regression tasks with \textit{E. coli} reservoirs \cite{ahavi_cellular_2025}, and even physically integrating conductive mycelium networks directly onto hardware as functional reservoir computers \cite{telhan_morphologically_2025}.
This combination could potentially simplify complex machine-learning models, and the hardware to operate them, in interactive systems. This would be particularly compelling for wearables and portable devices, which can be constrained by size and power.

\subsubsection{Biological Adaptation x Digital Memory}
Yet another identified gap is biological adaptation in combination with digital memory or state management. While some systems, such as \textit{Beauty} \cite{diblasi_beauty_2023}, utilize organisms' adaptive features, they lack digital memory support, missing the opportunity to create learning systems capable of accumulating knowledge over time, whether about the organisms themselves, their environments, or their interactions. Exploring this pathway could give rise to bio-digital systems that do more than react -- they could learn and co-evolve. Such future systems could serve as both a computational tool -- using bacterial adaptation as a component of problem-solving or machine learning -- and research platforms for understanding biological evolutionary processes. Digital state management would enable the system to ``remember'' successful adaptations, potentially guiding the recreation of beneficial bacterial configurations or inform applications in species preservation and environmental remediation.

% \new{Integrating biological adaptation with digital memory could thus enable reciprocal learning interactions that deepen over long durations, allowing user actions to produce lasting changes in biological behaviour while digital systems track, stabilize, or structure these changes, directly to form genuinely co-learning systems.}\ks{this is ok, but I think it's a repetition of what's already here so probably not needed}
% \ks{actually yes let's take this out...it's quite vague}

% \zb{maybe something from this example??? ---- a microbial consortia capable of integrating chemical and electronic signals to recognize patterns and learn through cellular memory. Platform built from  living cells by linking microbial sensing and communication with electronic networks. Since these microorganisms naturally communicate with each other through chemical or electrical signals, they can be linked to form a parallel computing system. Continuous culture systems will maintain microbial activity while allowing electronic interfacing to refine responses over time. This will allow the microbial networks to learn and adapt over time, enabling them to recognize patterns. This approach could enable computing systems to respond to real-world chemical inputs in ways that traditional hardware cannot. If successful, the project could advance medical diagnostics, environmental monitoring, and next-generation computing. -- https://news.rice.edu/news/2025/rice-research-team-quest-engineer-computing-systems-living-cells}

% \ek{perhaps we can find an example from Regenerative farms etc.?} 

\subsubsection{Temporal Diversification}
Our analysis also identifies opportunities to harness biological computation across broader temporal scales and patterns in bio-digital systems, taking into account both intrinsic and expressive temporal qualities. For example, fungal networks and slime moulds exhibit electrical spike trains \cite{whiting_slime_2014,miranda_composing_2018,mishra_sensorimotor_2024,adamatzky_reactive_2021,castellanos_microbial_2017,roberts_mining_2022,sedbon_c_2022}. These mirror the spiking patterns that neuromorphic hardware architectures, which promise low-power, low-latency solutions for edge intelligent computing, were specifically developed to emulate \cite{nawrocki_mini_2016,upreti_advancing_2025}. Instead of using silicon circuits to emulate biological spiking, neuromorphic processors could directly interface with biological spiking as input, potentially achieving even lower power consumption than conventional processors while operating within the organism's natural timing. 

The computational potential of biological processes across their layered temporal scales remains similarly unexplored. While their intrinsic metabolic processes occur at very high speeds, organisms' expressive activities -- ones that designers and users usually encounter and engage with in existing living artefacts (e.g., \textit{Bio-Digital Calendar} \cite{bell_bio-digital_2024}, \textit{Algae Relay} \cite{ikeya_designing_2023}) -- unfold much more slowly. Here, digital components can support the navigation of different temporal layers in both designing with and interacting with living materials. Take bacterial cellulose (BC), commonly used living materials in design and bio-HCI (e.g., \cite{bell_bio-digital_2024,krieger_living_2025, groutars_habitabilities_2025}), as an example. BC exhibits complex adaptations that are expressed over weeks to months. As a BC biofilm grows, it develops layers and structural variations that reflect its history, thickening in response to nutrient concentrations, changing textures with temperature fluctuations, etc. These slow changes could serve as biological memory, where the biofilm's physical structure encodes information (potentially spatially localized) about past environmental conditions and adaptation. 
Such layered computation enables new forms of cross-temporal interaction, in which users engage simultaneously with fast biological signals and slow material change. This resonates with prior HCI interest in temporal design \cite{oogjes_temporal_2024}, rhythm-based interaction \cite{costello_paying_2021}, and interfaces that integrate fast and slow feedback loops \cite{odom_extending_2021}.
% \ks{"addressing" didn't feel right, but I'm not sure about "echoing" either.}

%A system could harness this memory by using sensors to monitor these multi-dimensional growth patterns over extended periods. Digital components could provide evaluation by quantifying these responses over time, routing through directing specific nutrients or environmental conditions to different regions of the biofilm to create spatially distributed memory storage, growth, and adaptation, all with minimal power requirements.\ek{This is actually done in bioreactors for agitated fermentation but not for static. I would remove this (see my previous comment). distributed memory storage and co-adaptation is novel. }

%These slower processes, such as growth, adaptation, and evolution, represent what designers and users typically encounter and interact with in bio-digital systems.

% While the long, expressive timescales of processes such as growth, adaptation, and evolution are celebrated in regenerative and more-than-human design philosophies, they are computationally underutilized in current bio-digital systems. Embracing such timescales provides pathways for future bio-digital systems that not only honour biology's expressive timescales and foster appreciation for living organisms towards regenerative design goals but also make use of organisms' expanded computational functions, aligning with computational aspirations for sustainable and feature-rich paradigms.

\subsubsection{Spatial Diversification and Multiplicity}\label{sec:spatial}
% unexplored: only one organism 
With only 12\% of the systems we analysed operating at the ecosystem scale, bio-digital systems involving multiple organisms across different spatial scales represent yet more underexplored territory. Multiplicity -- the diversity of species in a living artefact \cite{groutars_designing_2024} -- offers unique opportunities for bio-digital systems. Digital capabilities can orchestrate inter-organism triggering mechanisms (or simply embrace naturally-occurring ones), creating cascading computational processes across biological networks that elevate the organisms involved, potentially benefitting the surrounding ecosystem and/or accomplishing tasks difficult to implement in silicon. For instance, one organism may secrete specific chemicals in response to environmental stimuli, activating a second organism to produce a more favourable growth environment for a third organism. Biologists have identified many suitable natural synergies, such as the collective relationships within lichens, or among \textit{Trichoderma} fungi, \textit{Bacillus} bacteria, and \textit{Pseudomonas} bacteria in agricultural biocontrol \cite{poveda_combined_2022}. By designing computational systems centred around these established biological partnerships, we could create interfaces that harness and highlight millions of years of co-evolutionary optimization between species, where cultures function together.

Foregrounding multiplicity and inter-species connections when designing bio-digital systems is especially critical for regenerative outcomes. High multiplicity systems, whether cultivated or wild, blur the boundary between the system and its surroundings. In such configurations, pollutants become power for the system, species within the system sustain one another, and the system generates resources that enrich ecosystems instead of depleting them. These inversions could transform extractive bio-digital relationships into mutualistic ones.

Relatedly, as Section \ref{sec:patterns} notes, the diversity in physical scale of our dataset was also quite limited, with most organisms confined to small containers in bio-digital systems. Expanding physical footprints, alongside multiplicity, would leverage microorganisms' scalability and open the door for bio-digital systems on architectural or landscape scales. Thus, we propose that future research should explore bio-digital systems at larger ecological and physical scales, where biological and digital components collaborate in ecosystem restoration, remediation, and regeneration. Such an agenda for HCI would extend current work from small, singular applications towards infrastructure-scale, ecological interventions that address urgent climate challenges.

% \begin{figure}[t]
%     \centering
%     \includegraphics[width=9cm]{figures/multi_organism_final.jpeg}
%     \caption{Illustrative concept of a bio-digital system using multipleorganisms. Three separate compartments, each containing a different species, are interconnected with one another and linked to digital instruments. Cascading interactions across species, where the activity of one organism triggers responses in another, perform complex computations.} %\zb{maybe this part is not necessary}
%     \Description{Line drawing of a conceptual bio-digital system. It shows three octagonal compartments each connected by thick cables to control panels and a screen featuring dials, switches, and digital displays, mounted on a shared frame. The compartments and digital elements are linked together, suggesting interactions between the organisms and computation.}
%     \label{fig:multi-species}
% \end{figure}

\subsection{Implications for Regenerative Ecologies} \label{sec:discussion}
%\ks{added comment from Elvin: "we could reflect on the taxonomy levels through the lens of regeneration, and consider whether they might be renamed or their definitions expanded further. (With a new table). Or we could keep the existing levels and table, but add guiding questions alongside the levels- from an ecological lens." (6 Sept: not sure I know what this means (anymore)...still relevant, or no?)}
%\ale{In terms of regeneration (as defined by Lyle, 1994 - I copied the definition in the latex), in addition to the point about the uni-directionality, we could also say something about the implications of combining potentially re-generative materials (the bio) with something that fundamentally isn't (the digital). Or maybe speculate on how the ability of the bio to self-replicate, adapat, resuscitate, etc. could open new possibilities for sustainable computing }\ks{very very subtly worked in the "fundamentally isn't" part...sustainable computing already touched upon in the next section... ;)}
%"Rooted in a living system approach, regenerative thinking in design suggests a profound understanding of living organisms, encompassing both human and non-human entities, and the ecologies they inhabit, to create human systems that can coevolve with natural systems, replenishing their inherent capacity to endure, flourish, and regenerate without depleting the essential life support systems and resources they rely on"

Achieving living artefacts for regenerative ecologies remains a challenging endeavour \cite{nicenboim_regenerative_2025}. Still, as we have discussed, exploring the diverse computational potentials of biological and digital components can already lead to richer regenerative possibilities. Our computational taxonomy reveals a misalignment between regenerative design philosophy -- which demands grounding in care and respect for the living systems they engage with \cite{karana_living_2023} -- and current bio-digital implementations. While regenerative frameworks advocate for systems that serve a broader ecological context, we have demonstrated that most existing microorganism-based bio-digital systems constrain organisms to unidirectional signal transduction or input/output roles, seemingly ``in service'' of their digital counterparts. At the same time, we also found that digital components, despite their vast potential computational functionalities, are currently also utilized merely as inputs to activate an organism, or as outputs to translate an organism's activity into something more perceivable to humans.

We propose that instead of defaulting to asking what computation advantages organisms might provide ``for'' digital systems, we must invert this perspective and systematically ask how digital components can computationally support organisms. This is a question that our taxonomy and subsequent analysis help to address. For instance, digital memory could preserve successful biological adaptations across generations; digital routing could optimize resource distribution in multi-species communities; digital evaluation could identify and amplify beneficial mutations. As previously discussed (Section \ref{sec:spatial}), uplifting multiplicity is a pathway that is especially fruitful for enriching ecological vitality.

On a more fundamental level, analysing bio-digital systems with our taxonomy also raises profound questions about what it means to use digital systems -- fundamentally non-living entities -- to provide computational support for living organisms and regenerative outcomes in the first place. For instance, does providing digital memory for bacterial adaptation, or activating an organism's response, actually constitute regeneration, or is this merely domestication? By designing multi-species interactions, who (or what) actually benefits: organisms, the ecosystem, the digital system, or human observers? While our taxonomy and analysis do not answer such questions, they surface them and provide scaffolding to begin to structurally address them.

\subsection{Implications for Human-Centred Computation}
Even for computational systems designers for whom regenerative outcomes are admittedly an afterthought, fuller bio-digital integration offers compelling advantages for human-centred systems. As mentioned in the introduction, achieving ubiquitous computing's vision of technology seamlessly blending into the environment requires moving from camouflaged silicon chips towards computation literally occurring through the living matter around us \cite{weiser_computer_1999}. In addition, organisms offer unique computational functions -- self-repair, growth, adaptation, and battery-free metabolic processes, to name a few -- that are impossible or prohibitively complex to implement digitally. For example, as previously discussed, DNA is an extremely information-dense, self-replicating storage medium that could help overcome bottlenecks in storage miniaturization. Moreover, implementing edge AI requires miniaturizing energy-hungry neural networks onto resource-constrained devices. If bacteria could perform evaluation, routing, and/or adaptation locally, perhaps digital components need only to handle simple input/output and communication tasks, making edge computing feasible with basic microcontrollers instead of specialized AI chips. MFCs harnessing bacteria's abilities to not only provide power but also sense and process information could enable drastic increases in operation time, without batteries or complex power management.
%MFCs that simultaneously sense and generate power have already been demonstrated, but further harnessing bacteria's abilities to process information could enable drastic increases in operation time, without batteries or complex power management.\ks{switch to this version after R&R...there is a forthcoming UIST'25 paper about using SMFCs for sensing that we will have to cite once it's public}

Additionally, our taxonomy-guided analysis reveals translatable theoretical insights and opportunities for computation-centric communities within HCI. The questions that surfaced during our analysis about what it means for computation to be ``for'' organisms mirror ongoing discussions in the human-AI interaction community around intention and explainability \cite{wang_designing_2019}, entanglement \cite{kuijer_co-performance_2018,frauenberger_entanglement_2019}, and uncertainty \cite{yang_re-examining_2020}. The emergence of codes such as ``output (connect to humans)'' reflects similar complications in crowd computing, where the role of digital platforms as mediators between requesters and workers raises questions about whose interests the computational system ultimately serves \cite{kittur_future_2013}. %\ks{prob not the right problem to reference...} \ale{Problem is OK, I think it can stay. I was thinking, in addition to mention things like a) modeling of computational properties of bio-stuff, mimicking the attempts to create programming languages for human computation (e.g., CrowdLang, TurKit). b) moving beyond ad-hoc interfaces between bio-digital and creating higher level abstractions that, for instance, would allow programming the interaction/communication. Something like a stochastic hybrid automaton.}\ks{I cut out the old meta-speculative stuff and added new meta-speculative stuff based on what I think this means} 
Our temporal coding challenges, such as distinguishing expressive and innate mechanistic timescales, also parallel tensions in human-AI collaboration, where humans and AI operate on different timescales. Just as Computer-Supported Cooperative Work (CSCW) frameworks account for asynchronous and synchronous communication \cite{rodden_survey_1991}, computational taxonomies for bio-digital and human-AI systems alike must account for simultaneous temporalities and their entanglements with agency.

Beyond these parallels, extending our computational understanding of living organisms opens possibilities for entirely new forms of programming and interaction. Mimicking attempts to create programming languages for human computation, such as \textit{CrowdLang} \cite{minder_how_2012} and \textit{TurKit} \cite{little_turkit_2010}, we could develop high-level abstractions capturing uniquely biological forms of computation, enabling entirely new programming languages that unlock new modes of interaction and control for bio-digital systems. This links to ongoing work on theoretical models for stochastic hybrid automata in formal methods and systems engineering, which similarly deal with complex systems exhibiting both deterministic and stochastic behaviours \cite{castaneda_stochastic_2011,hahn_compositional_2013}. There are many opportunities for cross-pollination here, with systematic bio-digital design practice informing conceptual models for broader HCI challenges and emerging forms of computational science.

%Perhaps the progressive temporal patterns prevalent among organisms can offer models for gradual trust building in human-AI relationships. Or perhaps building systems that embrace the sometimes lossy nature of biological memory (e.g., DNA is unstable at elevated temperatures) can inspire novel memory architectures that challenge the notion of perfect recall. \ks{thinking of deleting this second paragraph unless we come up w/ more concrete points (/if @Alessandro has anything)} \ale{See above. This last part is really meta-speculative - altough true. People are creating error detecting codes specific for dna }

% \newpage
%\subsection{Computational Taxonomy as a Shared Design Language}
\subsection{Shared Design Language as a First Step for Synergistic Bio-Digital Systems}
With our taxonomy we aimed to provide a vocabulary that is meaningful to both biology and computation. This bridge already proved fruitful during our own team's discussions. Computational opportunities identified through the taxonomy helped biodesigners recognize previously overlooked computational capabilities in organisms. Likewise, the taxonomy provided the vocabulary for computer scientists to understand the affordances of living organisms and subsequently propose concrete integration architectures.

Admittedly, the creation of this shared language and generation of opportunities are still very early steps towards synergistic bio-digital systems. While our taxonomy is designed to be tractable in existing hardware implementations, such as digital, analogue, neuromorphic, and cellular automata computing, the next step of implementing the previously discussed opportunities (and others) in practice faces its own technical hurdles. For instance, for DNA-based storage devices, maintaining DNA stability across temperature ranges, handling error correction, and designing reliable electronic interconnects with fluidic media are implementation challenges that need to be confronted. From the biological perspective, living media are inherently unpredictable. Even with years of experience working with the same organism, their rapid evolution and adaptation makes it challenging to design reliable systems. These difficulties become even more pronounced when attempting to scale up such systems. Overcoming these demands continued collaboration across multi-disciplinary teams like ours that integrate diverse perspectives, techniques, and methods.

Here, we would be remiss not to acknowledge the increasingly prominent role of generative AI (GenAI) in research. To that end, we believe that GenAI can offer additional support for tackling the upcoming implementation hurdles, and we envision our taxonomy continuing to be valuable there as well. While GenAI (via Large Language Models) has aided the discovery of new analogue circuit topologies \cite{gao_analoggenie_2025}, materials \cite{fuhr_deep_2022}, pharmaceuticals \cite{zeng_deep_2022}, and more, users struggle to formulate effective prompts that yield useful, actionable responses, frequently defaulting to opportunistic rather than systematic approaches \cite{zamfirescu-pereira_why_2023}. 

This challenge is currently acute in bio-digital system design. GenAI-supported tools like \textit{BioSpark}, which aims to bridge biological mechanisms (e.g., ``adaptive shape-shifting mechanism in intertidal microalgae'') and seemingly unrelated design goals (e.g., ``bike racks for sedans''), generate inspiring ideas but often suggest implementations with missing or incorrect details \cite{kang_biospark_2024}. Without a shared vocabulary between biodesigners and systems designers, we risk at least two types of problematic prompts: overly broad requests (``design a microbial computer'') or overly narrow component-focused instructions (``design a microbial system to act as a logic gate'').

Our taxonomy and generated opportunities enable more actionable, implementable prompting. For example, prompting Claude Sonnet 4 with ``design a system that uses biological material (organisms or DNA) for memory and digital components for routing/selection'' leads to a concrete implementation suggestion of using DNA within microfluidic chips with addressable channels, optical sensors for reading fluorescent markers, precision pumps, and other specific components. In contrast, the more naive prompt of ``design a storage system using biological material (organisms or DNA)'' leads to vague suggestions such as ``convert digital data to DNA sequences'' and ``retrieve by sequencing bacterial DNA.'' In this way, researchers can use our taxonomy as a guide for eliciting more useful suggestions from GenAI agents as they work towards implementing synergistic bio-digital systems in practice.

\subsection{Limitations and Future Work}\label{sec:future}
%\subsubsection{Extending Our Taxonomy}
The taxonomy we present in this paper opens pathways towards synergistic bio-digital design, providing vocabulary for conversations previously impossible. As more existing systems are analysed and new systems are designed, we anticipate the taxonomy will evolve as well, continuing to provide scaffolding for increasingly sophisticated computational partnerships between biological and digital agents. Importantly, we envision this refinement as a collective effort. We see our database as an organic resource that will expand through additional contributions from the wider community. Through future user studies, we will evaluate and iterate upon our database's structure and visualization platform in order to maximize their utility for biodesigners and computational systems designers alike.

We acknowledge that the taxonomy's implicit reliance on computational analogies may still constrain exploration of uniquely biological information processing mechanisms that have no digital equivalent. Future work could consider how emergent computational phenomena, such as morphological computation, swarm intelligence, or ecosystem-level information processing fit into, extend, or complicate our taxonomy. These might require embedding our functional descriptions into a multi-layered structure. Already, our coding process surfaced a distinction between living organisms' intrinsic and expressive timescales that would certainly be candidates for secondary layers in the temporal dimension. Additionally, our analysis focused on microorganisms and DNA, but the taxonomy's applicability to other systems, notably plant-based ones, remains a fruitful direction for future exploration. Plants introduce considerations, such as seasonal variations and larger physical scales, that will certainly inform additional taxonomic extensions. Furthermore, the taxonomy currently lacks dimensions for encoding organismal agency or consent, among other ethical considerations. Future iterations should incorporate frameworks for assessing power dynamics, reciprocity, and the distribution of (computational) benefits among digital, human, biological, and broadly other-than-human agents.

%\subsubsection{Beyond the Bio–Digital Dichotomy}
In this paper, for pragmatic reasons, we have analysed the biological and the digital separately. Conceptually, however, their individual interweaving calls for another orientation: one that moves beyond the biological-versus-digital dichotomy and instead embraces their underlying entanglement \cite{barad_meeting_2007,haraway_when_2007}. As the landscape of bio-digital systems grows, we envision perception, memory, communication, and even growth unfolding across both living and computational substrates in ways that render separation impossible. As biological and digital elements evolve from complementing to co-constituting one another, future frameworks must similarly evolve to pave the way for yet richer, more synergistic bio-digital systems.

\section{Conclusion}
This paper contributes a computational design taxonomy and vocabulary to facilitate the design of bio-digital hybrid systems for synergetic outcomes. We applied this taxonomy to analyse \artefact bio-digital systems spanning HCI, engineering, and art. We released the results as an open-source interactive database that presents our data within a rich design space -- one that can organically evolve and expand over time. Our analysis highlights opportunities for bio-digital systems that more fully engage the affordances of organisms across species, timescales, and ecosystems, while also fostering richer, reciprocal relationships between biological and digital functions. These directions outline interfaces that embrace the pillars of regenerative design, inviting more ecologically attuned and mutually beneficial futures for humans, organisms, and technology.

\begin{acks}
We thank our colleagues at DREAM and KiND for their valuable feedback and support. We also give special thanks to John Breed for providing the illustrations included in this paper

\end{acks}

\bibliographystyle{ACM-Reference-Format}
\bibliography{references}

%%
%% Appendix
%TC:ignore
\appendix
\section{Applicability of Taxonomy to Conventional Computing} \label{app:tradcompute}

\subsection{Computing Operations}

A computational design taxonomy for bio-digital systems should also have tractability in conventional computing. In this Appendix section we demonstrate how our functional taxonomy can be used to describe fundamental operations for arithmetic, sorting, and searching, as described in introductory computer science texts \cite{knuth_donald_e_art_2011}, as well as a machine learning algorithm.

\textbf{Basic Arithmetic (Addition):}
Addition receives numerical operands (\textit{Input}) and converts them into binary representation for processing (\textit{Transduction}). The algorithm performs bit-level comparisons and carry detection (\textit{Evaluation}), maintaining partial sums and carry bits in processor registers (\textit{Memory}). The final sum is returned in the appropriate format (\textit{Transduction} and \textit{Output}).

\textbf{Sort:}
Sorting algorithms receive an unsorted data sequence (\textit{Input}), sometimes requiring elements to be converted into comparable representations (\textit{Transduction}). Elements are then repeatedly compared to determine their relative ordering (\textit{Evaluation}). The algorithm maintains the current array state and intermediate orderings (\textit{Memory}), directing elements to appropriate positions based on comparison results (\textit{Routing}). Some sorting algorithms adapt their strategy based on data characteristics, for example by selecting optimal pivot points in quicksort (\textit{Adaptation}). The process concludes with the fully sorted sequence (\textit{Output}).

\textbf{Search:}
Search operations begin with a target value and data structure (\textit{Input}), often requiring the target to be converted to a searchable format (\textit{Transduction}). The algorithm repeatedly compares the target with elements in the search space (\textit{Evaluation}) while tracking current position and search boundaries (\textit{Memory}). Adaptive search algorithms adjust their strategy based on intermediate results, for example by reducing the search space in binary search (\textit{Adaptation}). The search returns the target's location, if it exists (\textit{Output}).

\textbf{Supervised Learning (Gradient Descent):}
Learning begins with labeled training data (\textit{Input}) that undergoes pre-processing and normalization (\textit{Transduction}). The algorithm computes loss functions to quantify prediction accuracy (\textit{Evaluation}) and determines gradients for updating parameters (\textit{Routing}). Model weights and optimizer states are maintained throughout the process (\textit{Memory}), with continuous adjustments to parameters and learning rates (\textit{Adaptation}). The trained model generates predictions (\textit{Transduction} and \textit{Output}).

\subsection{Hardware Architectures}
The taxonomy can also be used to describe diverse hardware architectures.

\textbf{Von Neumann (Digital) Computer:}
Von Neumann machines, the dominant architecture of modern computers \cite{godfrey_computer_1993}, process inputs through peripheral devices (\textit{Input}). Analogue-to-Digital Converters transduce continuous signals into discrete digital representations (\textit{Transduction}). The control unit decodes instructions and translates memory addresses, coordinating data movement between components (\textit{Routing}). Arithmetic, logical, and comparison operations are performed within the Arithmetic and Logic Unit (ALU) (\textit{Evaluation}). A unified memory system stores both program instructions and data (\textit{Memory}). Processed results are delivered through output devices, and often Digital-to-Analog Converters that convert digital signals back to analogue forms (\textit{Transduction}), for actuators and displays (\textit{Output}). Programmable instruction sequences can modify their own execution patterns based on data, enabling machine learning algorithms to update weights, adjust parameters, and optimize performance through iterative computation (\textit{Adaptation}).

\textbf{Analogue Computer:}
Analogue computers receive continuous signals through input terminals and sensor interfaces (\textit{Input}). Impedance matching circuits and operational amplifiers condition these signals into appropriate voltages and currents (\textit{Transduction}). The system performs continuous mathematical operations through analogue computing elements (\textit{Evaluation}). Switch matrices and multiplexers direct analogue signals through configurable processing paths and feedback loops (\textit{Routing}), while capacitive storage elements, delay lines, and analog shift registers maintain temporary system state (\textit{Memory}). Results are delivered through output amplifiers and signal conditioning circuits that drive external loads and display devices (\textit{Output}). Variable gain amplifiers and adaptive filtering circuits allow the system to tune transfer functions based on real-time data (\textit{Adaptation}).

\textbf{Neuromorphic Computer:}
Neuromorphic systems \cite{nawrocki_mini_2016} receive analog signals through sensors (\textit{Input}). Address-Event Representation (AER) circuits encode continuous signals into temporal spike trains (\textit{Transduction}). Arrays of artificial neurons perform distributed threshold comparisons and temporal integration in parallel (\textit{Evaluation}). Spike-based communication protocols route information via axonal pathways and synaptic connections (\textit{Routing}), with synaptic weights and membrane potentials stored in specialized analog memory circuits and memristive devices (\textit{Memory}). Processed spike patterns are converted back to analog signals through integrate-and-fire output neurons and digital interfaces (\textit{Output}). The system learns through spike-timing-dependent plasticity (STDP) and homeostatic mechanisms that modify synaptic strengths and neural thresholds based on activities (\textit{Adaptation}).

\textbf{Cellular Automata:}
Cellular automata (CA) computers, first proposed by von Neumann in 1966 \cite{neumann_theory_1966} and since implemented in a small number of hardware systems \cite{margolus_cellular_1987, toffoli_tommaso_cellular_1987}, receive as input a specification of initial states across a grid (\textit{Input}). State encoding circuits convert this data into discrete representations for each ``cell'' (\textit{Transduction}). Each cell evaluates its current state with respect to its neighbours' according to predefined transition rules (\textit{Evaluation}). Boundary condition handlers manage information flow between adjacent cells and across grid edges (\textit{Routing}). Distributed memory elements store current cell states and intermediate computational results (\textit{Memory}). Pattern detectors and circuits for state extractions (\textit{Transduction}) output the final grid configuration (\textit{Output}). With no memory beyond current cell states (\textit{Memory}), the CA architecture is not adaptive, though future variants may incorporate rule modification mechanisms (\textit{Adaptation}).

\section{Sources} \label{app:artefactsources}
In addition to the ACM Digital Library and IEEE Xplore, we searched review papers, hardcopy books, and art portfolios for microorganism-based bio-digital systems to analyse. These additional sources are listed in Table \ref{tab:extra_sources}.
\begin{table}[b]
\caption{Source types, sources and references for analysed systems}
\label{tab:extra_sources}
\centering
\small
\begin{tabular}{p{1.6cm} p{5.7cm} p{1.2cm} r}
\toprule
 \textbf{Source Type} & \textbf{Source} & \textbf{Reference} \\
\midrule

 \multirow[t]{6}{*}{Review Papers} 
 & Grouters et al., 2024 & \cite{groutars_designing_2024} \\
 & Ikeya et al., 2025 & \cite{ikeya_aesthetics_2025} \\
 & Karana et al., 2020 & \cite{karana_living_2020} \\
 & Kim et al., 2023 & \cite{kim_surfacing_2023} \\
 & Pataranutaporn et al., 2020 & \cite{pataranutaporn_living_2020}\\
 & Zhou et al., 2022 & \cite{zhou_habitabilities_2022} \\
 \addlinespace[1em]
 \multirow[t]{6}{*}{Books}
 & \textit{Art and Electronic Media}, Edward A. Shanken, 2009 & \cite{edward_art_2009}\\
 & \textit{Art + Science Now}, Stephen Wilson, 2010 & \cite{wilson_art_2010} \\
 & \textit{Bio Art Altered Realities}, William Myers, 2015 & \cite{myers_bio_2015} \\
 & \textit{Biodesign}, William Myers, 2014 &  \cite{myers_bio_2018}\\
 & \textit{BioMedia}, Peter Weibel, 2023 & \cite{weibel_biomedia_2023} \\
 & \textit{Kunst en Wetenschap}, Peter de Jaeger, 2020 & \cite{jaeger_kunst_2020} \\
 
 \addlinespace[1em]
 \multirow[t]{6}{*}{Art Portfolios}
 & \url{http://allisonx.com} & \cite{kudla_allison_2025} \\
 & \url{http://ivanhenriques.com} & \cite{henriques_ivan_2025} \\
 & \url{http://interspecifics.cc} & \cite{interspecifics_interspecifics_2016} \\
 & \url{http://michaelsedbon.com} & \cite{sedbon_michael_2022} \\
 & \url{http://novainnova.com} & \cite{van_oers_nova_2025} \\
 & \url{http://teresavandongen.com} &  \cite{van_dongen_teresa_2025}\\
\bottomrule
\end{tabular}
\end{table}

% \newpage
\section{Excluded Systens} \label{app:excluded}
Examples of systems involving microorganisms (or DNA) that were screened out during Phase 2 of our analysis include the following:

\begin{itemize}
    \item Equipment used for biological studies or observations
    \item Optimizations (e.g., higher efficiency) of previously-reported microbial fuel cells or plant–microbial fuel cells
    \item Optimizations (e.g., higher efficiency) of previously-reported biosensors
    %\ks{what are these? like ones that make extensive use of micro/nanofab?} \zb{This is an example: "Evaluating Nonlinear Impedance Excitation as Detection Method for Biosensors" or "An Intelligent and Optimal Deep-Learning Approach in Sensor-Based Networks for Detecting Microbes" basically papers that is about the optimization of biosensor}\ks{new wording is clear}
    \item Systems developed solely for species identification or detection without further interaction (e.g., detection of pesticides or algae) 
    \item Systems developed as by-products of studies to optimize cultivation methods
    %\item Biosensors developed solely for measurement purposes, where organisms are used only as biological sensing elements \ks{what are these?} \zb{For example "Biosensor based on biochip-G for dissolved oxygen detection from photosynthesis process of green algae chlorella vulgaris these biosensors do only incorporate the incorporates the organism as the biological recognition element}
    \item Technologies aimed solely at removing organisms (e.g., algae removal)
    \item Organism-based sensors where digital components are absent or physically detached and only used for later result read-out
\end{itemize}

Articles that only take inspiration from microorganism but do not integrate actual organisms, e.g.:
\begin{itemize}
    \item Insect-inspired robots
    \item Augmented/virtual reality (AR/VR) systems for interacting with or observing organisms
    \item Organism-inspired algorithms
\end{itemize}

\onecolumn
\section{Included Bio-Digital Systems} \label{app:included}

{\small
\begin{longtable}{lp{4.4cm}p{7.8cm}l}
\caption{List of systems analysed for taxonomy development
\label{tab:artefact}}\\
\toprule
Year & Author/Artist & System Name/Title (Abbreviated) & Ref. \\
\midrule
\endfirsthead

\multicolumn{4}{c}{{\bfseries \tablename\ \thetable{} -- continued from previous page}} \\
\toprule
Year & Author/Artist & System Name/Title (Abbreviated) & Ref. \\
\midrule
\endhead

\midrule \multicolumn{4}{r}{{Continued on next page}} \\
\endfoot

\bottomrule
\endlastfoot

2005 & Ioannis Ieropoulos et al. & Ecobot-II & \cite{ieropoulos_ecobot-ii_2005}\\
2006 & Iosif Pinelis et al. & A Micro ``Flea Circus'' & \cite{pinelis_micro_2006}\\
2007 & Walder André \& Sylvain Martel & Autonomous Microrobot Propelled by Magnetotactic Bacteria & \cite{andre_preliminary_2007}\\
2007 & Jeffrey J. Tabor & Programming Living Cells as Massively Parallel Computers & \cite{tabor_programming_2007}\\
2008 & Adrian David Cheok et al. & Empathetic living media & \cite{cheok_empathetic_2008} \\
2008 & Allison Kudla & The Crucible & \cite{kudla_crucible_2008}\\
2010 & Douglas Easterly et al. & Tardigotchi & \cite{easterly_tardigotchi_2010}\\
2010 & Ingmar H. Riedel-Kruse et al. & Biotic Games & \cite{riedel-kruse_design_2010}\\
2011 & C-Lab & Stress-o-stat & \cite{c-lab_stress-o-stat_2011}\\
2011 & Mathijs Munnik & Microscopic Opera & \cite{munnik_microscopic_2011}\\
2011 & Philips Design Probe Group & Microbial Home & \cite{philips_design_microbial_2011}\\
2013 & C-Lab & Living Mirror & \cite{c-lab_living_2013}\\
2014 & Foad Hamidi \& Melanie Baljko & Rafigh & \cite{hamidi_rafigh_2014}\\
2014 & Ivan Henriques & Symbiotic Machine & \cite{henriques_symbiotic_2014}\\
2014 & James G.H. Whiting et al. & Slime Mould Logic Gates & \cite{whiting_slime_2014}\\
2015 & ecoLogicStudio & Urban Algae Canopy & \cite{ecologicstudio_urban_2015}\\
2015 & Interspecifics & GFP Screen & \cite{interspecifics_gfp_2015}\\
2015 & Seung Ah Lee et al. & Tangible Interactive Microbiology & \cite{lee_tangible_2015}\\
2015 & Seung Ah Lee et al. & Trap it! & \cite{lee_trap_2015}\\
2016 & Edward Braund et al. & Physarum-Based Memristors for Computer Music & \cite{adamatzky_physarum-based_2016}\\
2016 & EMW Street Bio & Biota Beats & \cite{emw_street_bio_biota_2016}\\
2016 & Lukas C Gerber et al. & BioGraphr & \cite{gerber_biographr_2016}\\
2016 & Ivan Henriques & Caravel & \cite{henriques_caravel_2016}\\
2016 & Interspecifics & Micro-Rhythms & \cite{interspecifics_micro-rhythms_2016}\\
2016 & Gènes-Hélène Isitan & Where Species Meet & \cite{isitan_where_2016}\\
2016 & Honesty Kim et al. & LudusScope & \cite{kim_ludusscope_2016}\\
2017 & Carlos Castellanos & Microbial Sonorities & \cite{castellanos_microbial_2017}\\
2017 & Gilberto Esparza & BioSoNot & \cite{esparza_biosonot_2017} \\
2017 & Marin Sawa et al. & Electricity generation from digitally printed cyanobacteria & \cite{sawa_electricity_2017} \\
2017 & Helene Steiner & Bixels DNA Tetris & \cite{steiner_bixels_2017}\\
% 2017 & Gilberto Esparza & BioSoNot & \cite{esparza_biosonot_2017}\\
2017 & Catalina Puello et al. & Living Screens & \cite{puello_living_2017}\\
2017 & Maurizio Rossi et al. & Let the Microbes Power Your Sensing Display & \cite{rossi_let_2017}\\
2017 & Maurizio Rossi et al. & Wireless Sensing Powered by Plant-Microbial Fuel Cell & \cite{rossi_long_2017}\\
2017 & Tom J. Zajdel et al. & Biohybrid Sensing Platform Built on Bacterial Flagellar Motor & \cite{zajdel_towards_2017}\\
2018 & Laura Grebenstein et al. & Biological Optical-to-Chemical Signal Conversion Interface & \cite{grebenstein_biological_2018}\\
2018 & Raphael Kim et al. & Mould Rush & \cite{kim_new_2018}\\
2018 & Eduardo Reck Miranda et al. & Composing with Biomemristors & \cite{miranda_composing_2018}\\
2019 & Tim Dobbs & BioArtBot & \cite{dobbs_bioartbot_2019}\\
2019 & Ivan Henriques & Bacterbrain & \cite{henriques_bacterbrain_2019}\\
2019 & Micheal Sedbon & CMD & \cite{sedbon_cmd_2019}\\
2020 & Rachel Armstrong and Julie Freeman et al. & ALICE & \cite{armstrong_active_2020}\\
2020 & Jacob Douenias et al. & Living Things & \cite{douenias_living_2020}\\
2020 & Amy T. Lam et al. & Pac-Euglena & \cite{lam_pac-euglena_2020}\\
2020 & Kyungwon Lee et al. & MicroAquarium & \cite{lee_microaquarium_2020}\\
2020 & Nova Innova & Living Water - POND & \cite{nova_innova_pond_2020}\\
2021 & Andrew Adamatzky et al. & Reactive Fungal Wearable & \cite{adamatzky_reactive_2021}\\
2021 & Bahareh Barati et al. & Living Light Interface & \cite{barati_living_2021} \\
2021 & Dominique Chen et al. & Nukabot & \cite{chen_nukabot_2021}\\
2022 & Gilberto Esparza & Bio-Electrólisis & \cite{esparza_bio-electrolisis_2022}\\
2022 & Kyungwon Lee et al. & EuglPollock & \cite{lee_euglpollock_2022}\\
2022 & Xiaomeng Liu et al. & Microbial Biofilms for Electricity Generation & \cite{liu_microbial_2022}\\
2022 & Jasmine Lu \& Pedro Lopes & Living Organisms in Devices for Care-based Interactions & \cite{lu_integrating_2022}\\
2022 & iGEM Paris Bettencourt & Bacterial Networked Electronic Interface & \cite{igem_paris_bettencourt_bacterial_2022}\\
2022 & Nic Roberts \& Andrew Adamatzky & Mining Logical Circuits in Fungi & \cite{roberts_mining_2022}\\
2022 & Micheal Sedbon & Ctrl & \cite{sedbon_c_2022}\\
2022 & Micheal Sedbon & Cryptographic Beings & \cite{sedbon_cryptographic_2022}\\
2022 & Where Dogs Run & Archaean Memory Farm & \cite{inozemtseva_archaean_2022} \\
2023 & Johnny DiBlasi et al. & Beauty & \cite{diblasi_beauty_2023}\\
2023 & Yuta Ikeya \& Bahar Barati & Algal Relay Computer & \cite{ikeya_designing_2023}\\
2023 & Su-Jin Song et al. & Navigated Biobot Swarms of Bacteria \textit{Magnetospirillum} & \cite{song_precisely_2023}\\
2023 & Jiwei Zhou et al. & Cyano-Chromic Interface & \cite{zhou_cyano-chromic_2023}\\
2024 & Fiona Bell et al. & Bio-Digital Calendar & \cite{bell_bio-digital_2024}\\
2024 & Ahmet Bilir \& Sema Dumanli & Biodegradable Implant Antenna with Genetically Modified Bacteria & \cite{bilir_biodegradable_2024}\\
2024 & Zoë Breed et al. & Algae Alight & \cite{breed_algae_2024}\\
2024 & Anand Kumar Mishra et al. & Robots Mediated by Electrophysiological Measurements of Mycelia & \cite{mishra_sensorimotor_2024}\\
2024 & Valentin Postl et al. & Mold Printer & \cite{postl_mold_2024}\\
2024 & Ðan Vy Vu et al. & Biofabrication with Mycelium & \cite{vu_addressing_2024}\\
2024 & Philip Wijesinghe & Light-Deformable Microrobots & \cite{wijesinghe_light-deformable_2024}\\
2025 & Mukrime Birgul Akolpoglu et al. & Navigating Microalgal Magnetic Biohybrids & \cite{akolpoglu_navigating_2025}\\
2025 & Mengyao Guo \& Jinda Han & Slimo City & \cite{guo_slimo_2025}\\
2025 & Alexandra Teixeira Riggs et al. & Mold Sounds & \cite{riggs_mold_2025}\\

\end{longtable}
}

\section{Resolving Ambiguities}
\label{app:abm_convo}

Example coding discussion between the two team members who coded the systems to resolve ambiguities in the project Mould Rush \cite{kim_new_2018}.

\begin{quote}
\textbf{Coder 1:} I was unsure how to classify the Mould Rush temporal patterns. I thought \textit{progressive} might apply, but perhaps also \textit{transient}? \textit{progressive}, because the mould keeps growing over time. But in the game, you can also kill the mould by dropping bleach on it. Does that make it \textit{transient}?

\textbf{Coder 2:} Let me open the article and read through the project details again. 

\textbf{Coder 2:} I agree with \textit{progressive}. I wouldn’t call it \textit{transient}, because even when killed, the mould is still physically present. You would have to clean it off. It’s just dead, not growing. So only \textit{progressive}.

\textbf{Coder 1:} Should we also include \textit{death} than as an observable output?

\textbf{Coder 2:} Hm, yes I would say both \textit{death} and \textit{growth} are observable outputs that the player can perceive.

\textbf{Coder 1:} Okay I will put that down. One last question about this project, this one was really difficult to code. For the trigger mechanism I would say \textit{chemical}, since they use bleach to kill the organism. But what about for the growth of the mould? The human player inserts the mould on the Petri dish does this count as a trigger?

\textbf{Coder 2:} For killing, definitely \textit{chemical}. For growth, I would say \textit{none}, because there no external stimuli to let them grow. This is just the autonomous
behaviours. So code both \textit{chemical} and \textit{none}.
\end{quote}

%TC:endignore

\end{document}